\documentclass[12pt]{iopart}

\usepackage{iopams}  
\usepackage{graphicx}
\usepackage{color}
\def\beq{\begin{equation}}
\def\eeq{\end{equation}}

\begin{document}
\title[]{Mobility gap and quantum transport in functionalized graphene bilayer} 

\author{
Ahmed {Missaoui}$^1$,
Jouda J. {Khabthani}$^2$,
Nejm-Eddine {Jaidane}$^1$,
Didier {Mayou}$^{3,4}$,
Guy {Trambly de Laissardi\`ere}$^5$
}

\address{$^1$ Laboratoire de Spectroscopie Atomique Mol\'eculaire et Applications, 
D\'epartement de Physique, Facult\'e des Sciences de Tunis, Universit\'e de Tunis El Manar,  Campus universitaire 1060 Tunis, Tunisia}
\address{$^2$ Laboratoire de Spectroscopie Atomique Mol\'eculaire et Applications, 
D\'epartement de Physique, Facult\'e des Sciences de Tunis, Universit\'e de Tunis El Manar,  Campus universitaire 1060 Tunis, Tunisia}
\address{$^3$ CNRS, Inst NEEL, F-38042 Grenoble, France.}
\address{$^4$ Universit\'e Grenoble Alpes, Inst NEEL, F-38042 Grenoble, France}
\address{$^5$ Laboratoire de Physique th\'eorique et Mod\'elisation, CNRS and Universit\'e de Cergy-Pontoise, 
95302 Cergy-Pontoise, France}
\ead{guy.trambly@u-cergy.fr}

\vspace{10pt}
\begin{indented}
\item[]March 2018
\end{indented}

\begin{abstract}
In a Bernal graphene bilayer, carbon atoms belong to two inequivalent sublattices A  and B, with  
atoms that are coupled to the other layer by  $p_\sigma$ bonds belonging to sublattice A  
and  the other atoms belonging to sublattice B. 
We analyze the density of states and the conductivity of Bernal graphene bilayers when atoms of sublattice A or B 
only are randomly functionalized. We find that for a  selective functionalization on sublattice B only,  a mobility gap of the order of  $0.5$\,eV is formed close to the Dirac energy  at concentration  of adatoms $c\geq 10^{-2}$. 
In addition, at some other energies conductivity presents anomalous behaviors. 
We show that these properties are related to the bipartite structure of the graphene layer.
\end{abstract}

%
\vspace{2pc}

%
%
%

\section{Introduction}

Electronic properties at nanoscale are the key to the novel applications of low-dimensional and nanomaterials 
in electronic and energy technologies. 
In particular, a lot of research has been devoted to understanding the remarkable electronic 
structure and transport properties of  bilayer (or multilayers) of graphene  
\cite{Novoselov04,Berger04,Hashimoto04,Zhang05,Wu07,Zhou08,Robinson08,Castro09_RevModPhys,Chen09,Bostwick09,Ni10,Nakaharai13,Zhao15,Barrejon15}. 
Depending on the stacking, 
the charge carriers were shown, both theoretically 
\cite{Ferreira11,McCann13,Trambly10,Trambly12,Trambly16,VanTuan16,Missaoui17} 
and experimentally \cite{Ohta06,Zhang09,Stabile15,Ulstrup14,Ju15,Rozhkov16}, 
to behave like massless Dirac particles or massive particles with chirality. 
Electronic 
properties can be tuned by various means and 
in particular 
by electrostatic gate or by adding of static defects and functionalization by adatoms or admolecules of monolayer (MLG) \cite{Robinson08,Lherbier08,Yuan10b,Leconte10,Skrypnyk10,Skrypnyk11,Leconte11,Lherbier11,Trambly11,Lherbier12,Trambly13,Trambly14,Gargiulo14,Yuan12,Zhao15,Barrejon15,Yan16,Xu16} and bilayer (BLG) \cite{Ju15,Stabile15,VanTuan16,Missaoui17}. 
For example, 
one can open a band gap in this system by electrostatic gating \cite{Ohta06,Zhang09}. Recently such locally coupled structures have been also observed in chemical vapor deposition (CVD) graphene samples \cite{Ju15,Yan16} where, due to rippling, the layers were decoupled in some regions, while being connected in others. It has also been shown that UV irradiation, which results in water dissociative adsoption on graphene of few \% of adsorbates, can induce a tunable reversible gap \cite{Xu16}. 

In this work, we investigate the density of states and the conductivity of a Bernal bilayer graphene (BLG) when the upper layer is functionalized by adatoms. There are two types of site on the upper layer, as shown in Figure  \ref{Fig_structure_biAB}. Sites of sublattice A  are above a carbon atom of the lower layer whereas sites of sublattice B are not.  Therefore it is possible in principle to functionalize selectively atoms which belong to  sublattice A or to sublattice B only. We consider here, that, within the functionalized sublattice, the repartition of the functionalized atoms is random. As a main result we find that, when only sublattice B is functionalized a mobility gap of the order of  $0.5$\,eV is formed close to the Dirac energy  at concentration  of adatoms $c\geq 10^{-2}$. Furthermore for both sublattice functionalization the conductivity increases in some Fermi energy window, when the concentration of functionalized sites increases. This is because the functionalization is not just introducing scattering centers but deeply changes the electronic structure.  As we show the creation of the gap and the abnormal behavior of the conductivity are related to the  bipartite nature of the monolayer and bilayer graphene. 

\section{Method}

\begin{figure}
\begin{center}

\includegraphics[width=7cm]{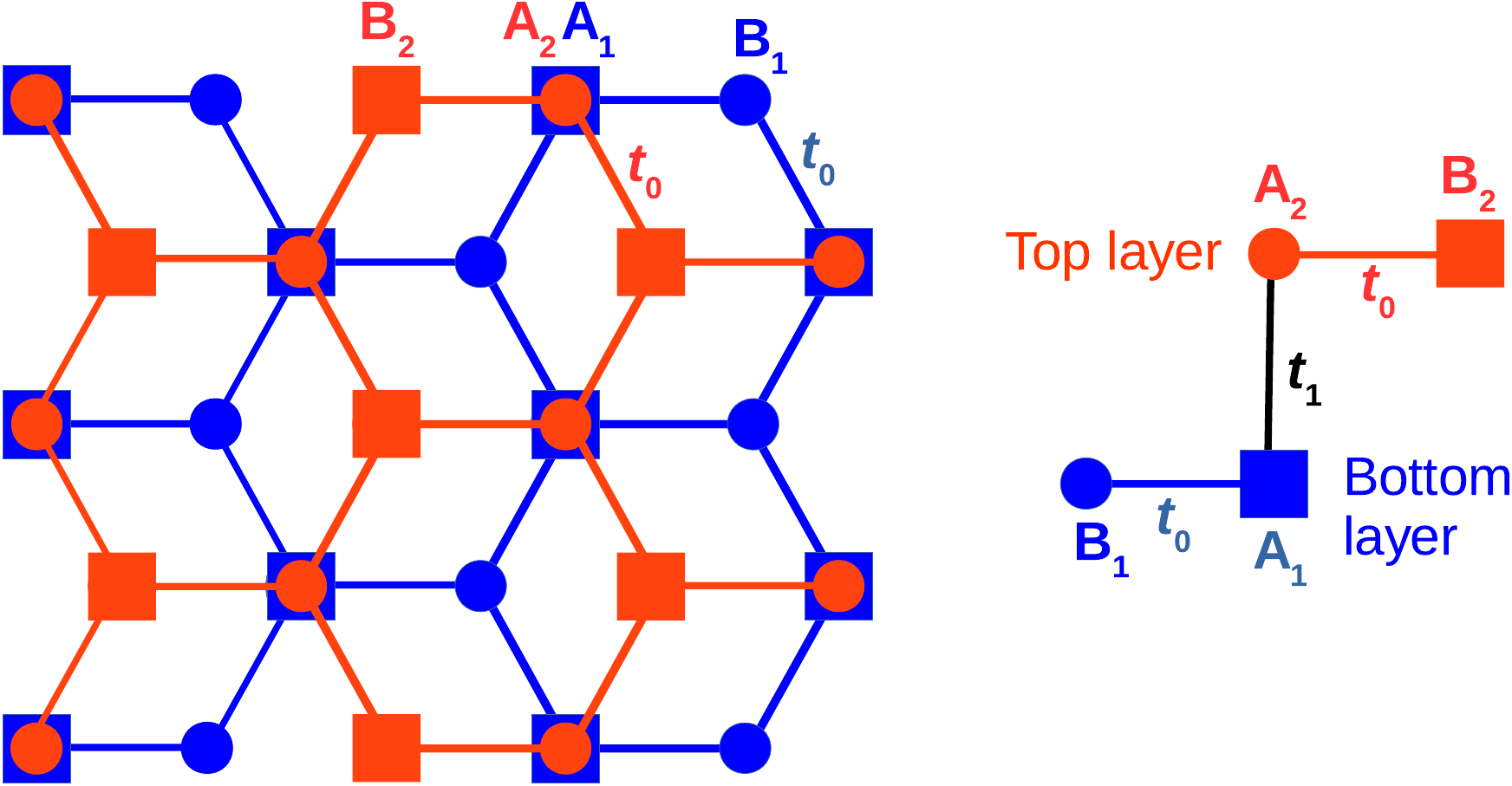}

\end{center}

\caption{ \label{Fig_structure_biAB}
Bilayer structure with sublattice $\alpha = \{ A_1 , B_2 \}$ (square), and  sublattice $\beta = \{ A_2 , B_1 \}$ (circle).
}
\end{figure}

The BLG studied here consist of the bottom layer 1 and of the top layer 2 as shown in Figure  \ref{Fig_structure_biAB}. The top layer 2 is functionalized whereas the bottom layer 1 keeps its perfect structure. There are four carbon atoms in the unit cell, two carbons A$_1$, B$_1$ in layer 1 and A$_2$, B$_2$ in layer 2 where A$_2$ lies on the top of A$_1$.
We use an electronic model where only $p_z$ orbitals are taken into account, since we are interested in the low energy physics i.e. electronic states close to the Dirac energy. The adsorbates which create a covalent bond with a carbon atom of the graphene upper layer  is represented by removing the $p_{z}$ orbitals of the functionalized carbon atoms \cite{Pereira06,Pereira08a,Robinson08,Wehling10,Skrypnyk10,Ducastelle13}.  
The missing $p_z$ orbitals are distributed randomly only on  sites of the top  layer 2 in the sublattice A or B. 
The tight-binding (TB) Hamiltonian for $p_z$ orbitals has the form:
\beq
\hat{H} = \sum_{\langle i,j \rangle} t_{ij} \left( c_i^{\dag}c_j + c_j^{\dag}c_i \right)
\label{eq_TB}
\eeq  
where $c_i^{\dag}$ and $c_i$ create and annihilate respectively an electron on site $i$,
$\langle i,j\rangle$ is  the sum on index $i$ and $j$ with $i\ne j$, 
and  $t_{ij}$ is the hopping matrix element
between two $p_z$ orbitals $i$ and $j$.
We analyze the average local density of states (LDOS) 
on the sublattices  A or B  of each plane, and the conductivity as a function of the position of the Fermi energy. 
Densities of states are computed by recursion (Lanczos algorithm) \cite{methodeRecursion} in real-space on sample containing a few $10^{7}$ carbon atoms with periodic boundary conditions. 
Within the Kubo-Greenwood formalism we compute the microscopic conductivity $\sigma_m(E)$ \cite{Trambly13} using the real-space method 
developped by Mayou, Khanna, Roche and Triozon \cite{Mayou88,Mayou95,Roche97,Roche99,Triozon02}
(see supplementary material \cite{supplementaryMat} Sect. 4). 
$\sigma_m$ is the semi-classical conductivity that does not take into account the quantum corrections due to multiple scattering effects. Typically this quantity represents a room temperature conductivity when multiple scattering effects are destroyed by dephasing due to  the electron-phonon scattering.

\section{Results}

We present first calculations performed  with the standard nearest neighbor hopping Hamiltonian  (TB1): $t_0=2.7$\,eV for intra-layer hopping between A and B atoms, and  $t_1=0.34$\,eV for  nearest neighbor inter-layer hopping between A$_1$ and A$_2$ atoms. 
The advantage of this simple Hamiltonian TB1 is to allow a detailed physical discussion of the physical mechanism involved. These  results are confirmed  by analyzing a more realistic Hamiltonian description that takes into account hopping beyond the nearest neighbor hopping model (TB2) (supplementary material \cite{supplementaryMat}, Sect. 1). 
TB2 has been used successfully to study the electronic structure in rotated bilayer of graphene \cite{Trambly10,Trambly12,Trambly16} in good agreement with STM density of states measurements \cite{Brihuega12,Cherkez15} and for transport calculations \cite{Trambly11,Trambly13,Trambly14,Missaoui17}.

\subsection{Results with nearest-neighbor hopping  Hamiltonian (TB1)}

\begin{figure}
\begin{center}

\includegraphics[width=4.1cm]{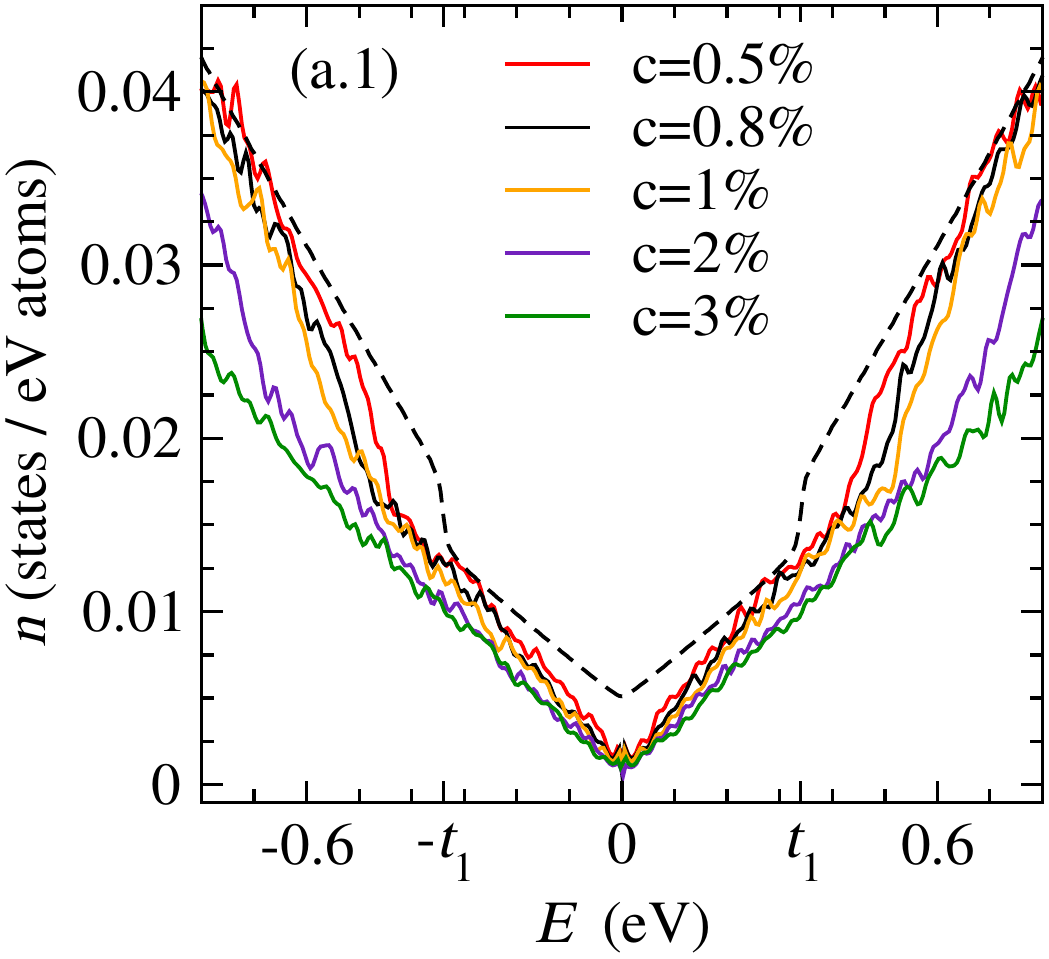} 
~ \includegraphics[width=4.1cm]{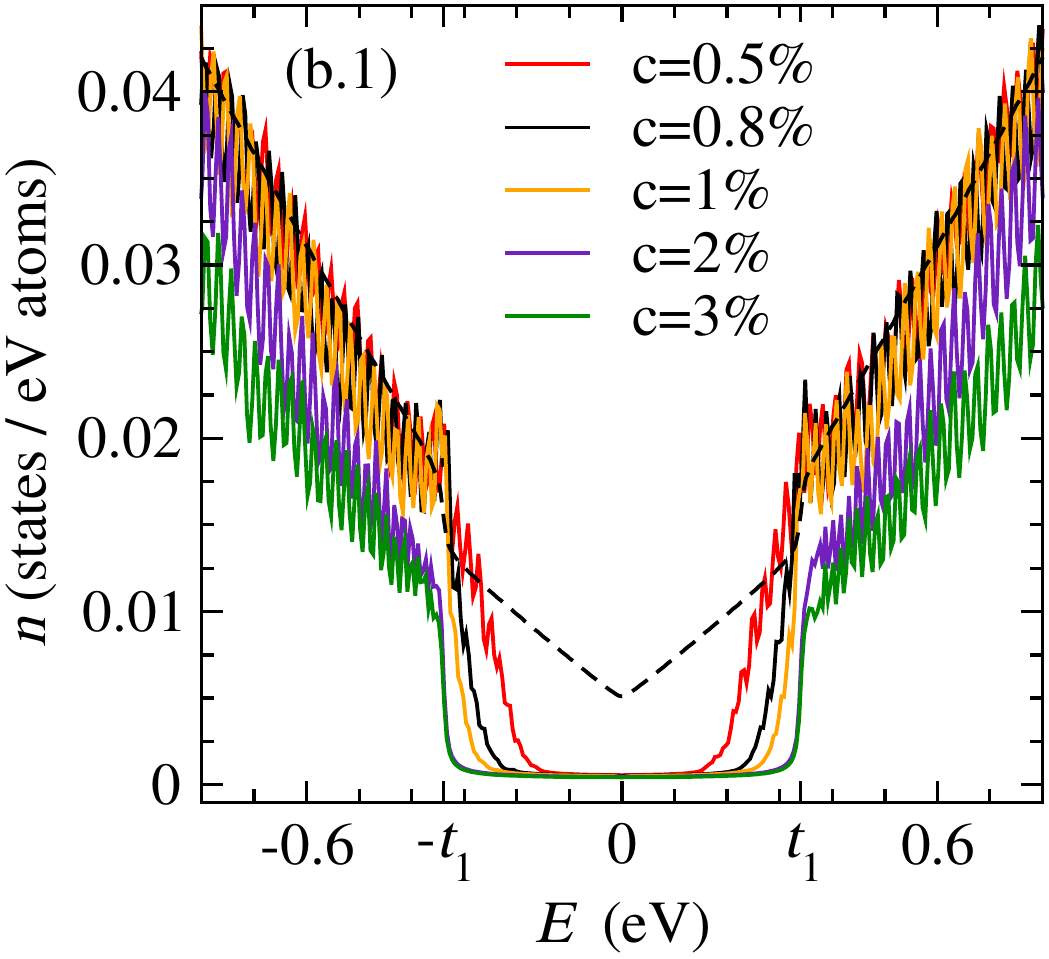}~ 

\vspace{.1cm}
\includegraphics[width=4.2cm]{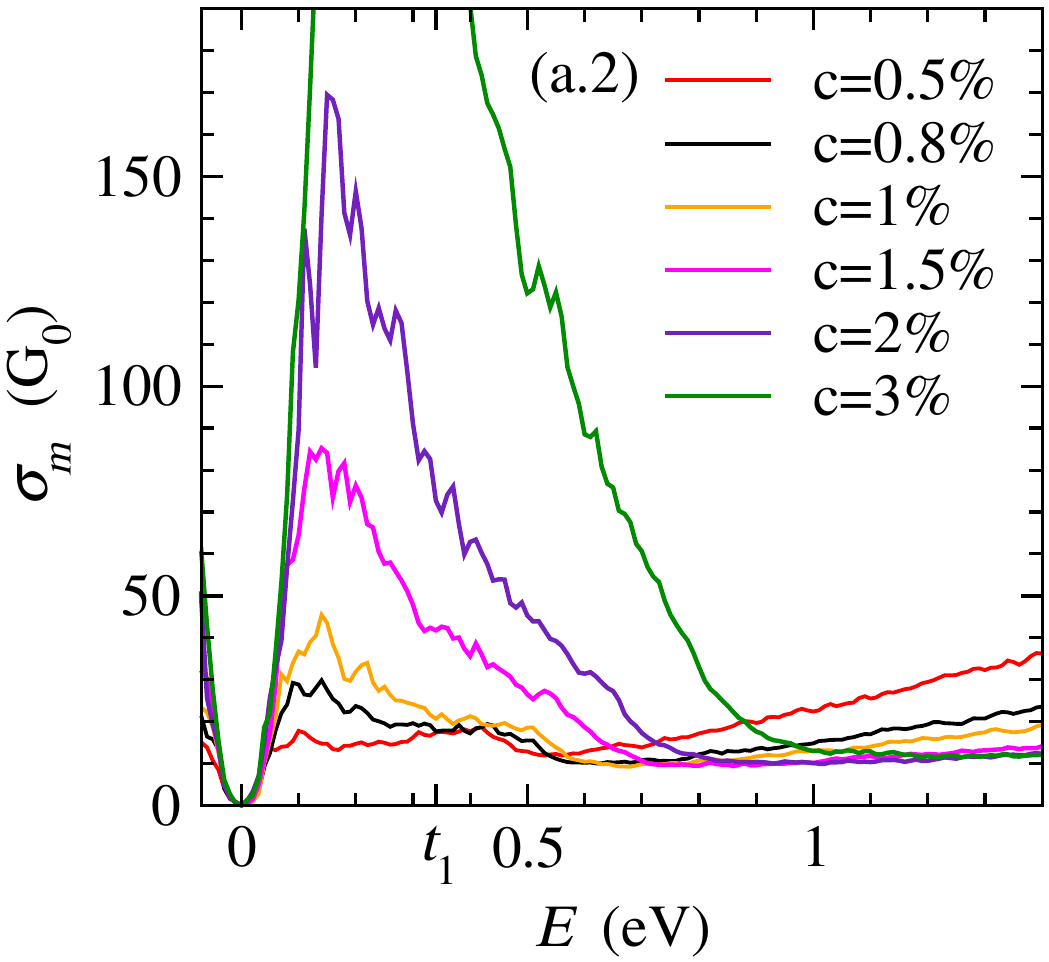} 
~ \includegraphics[width=4.1cm]{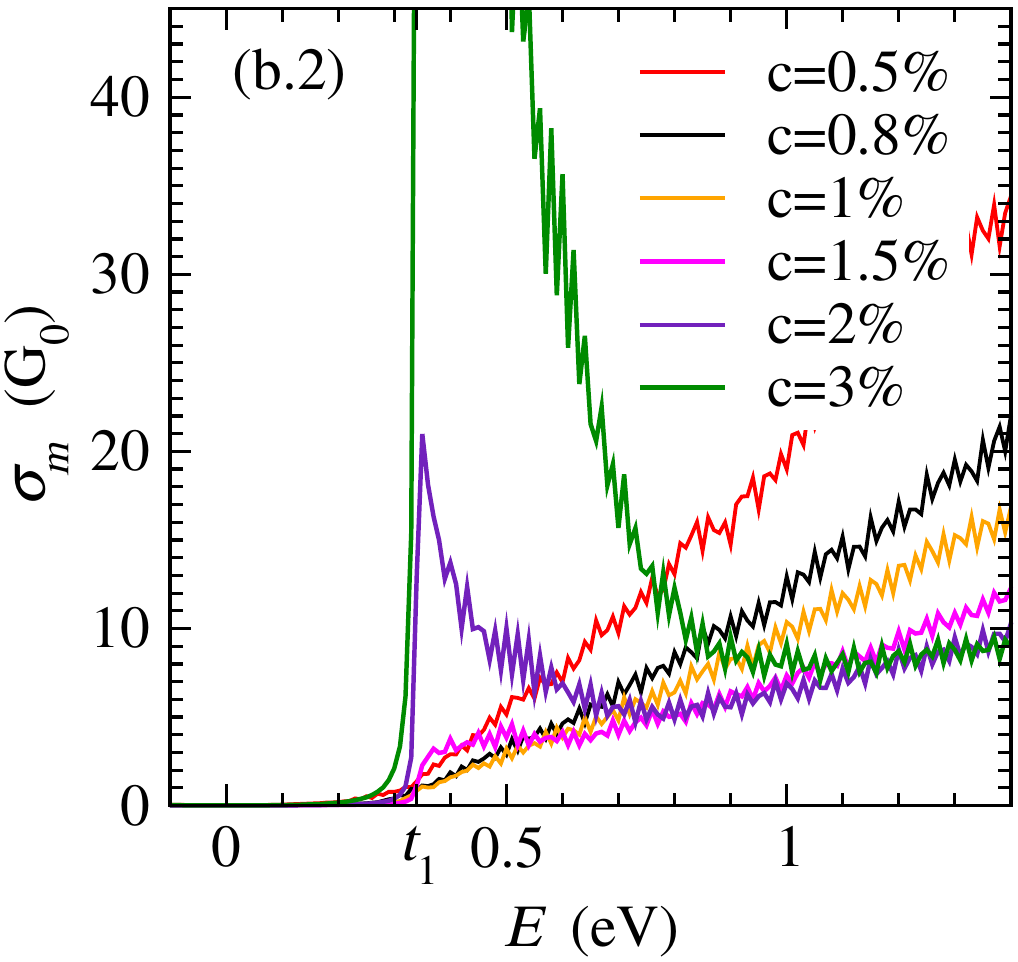} 

\vspace{.1cm}
\includegraphics[width=4.1cm]{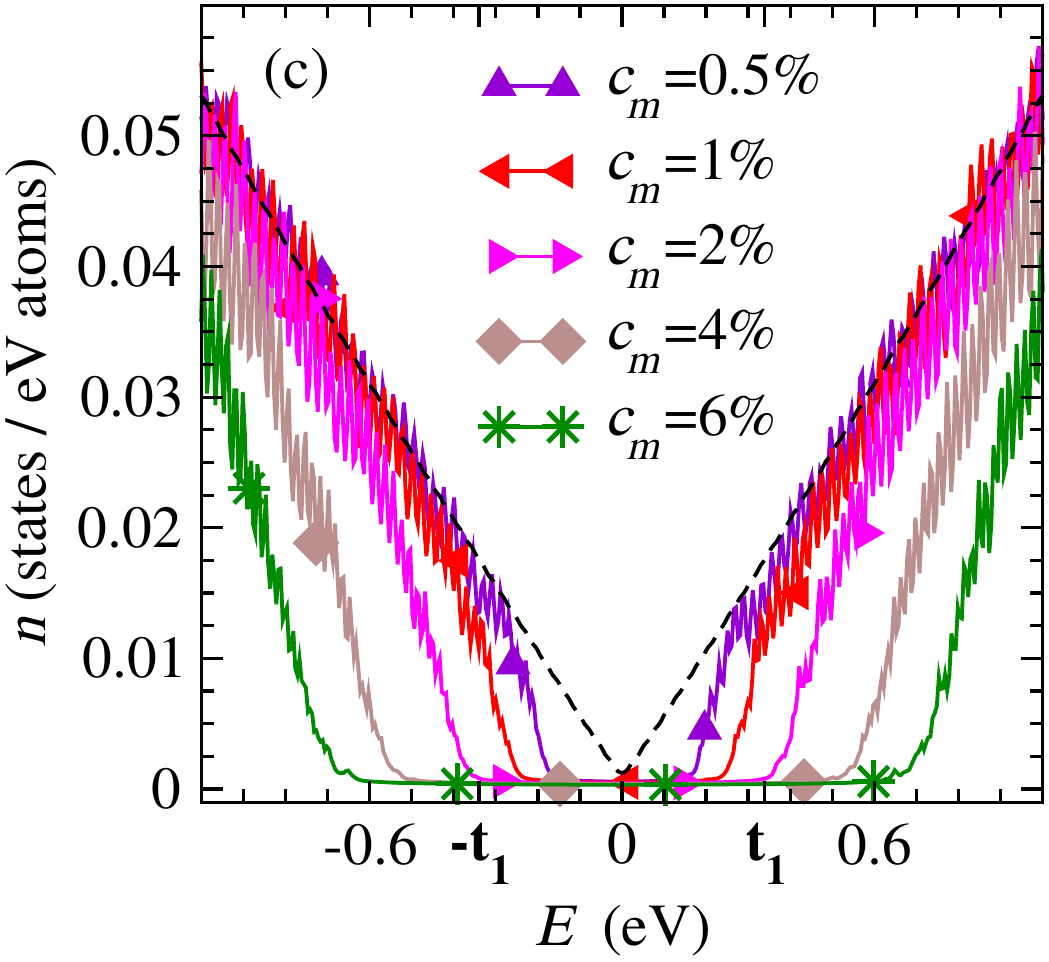} 
~ \includegraphics[width=4.1cm]{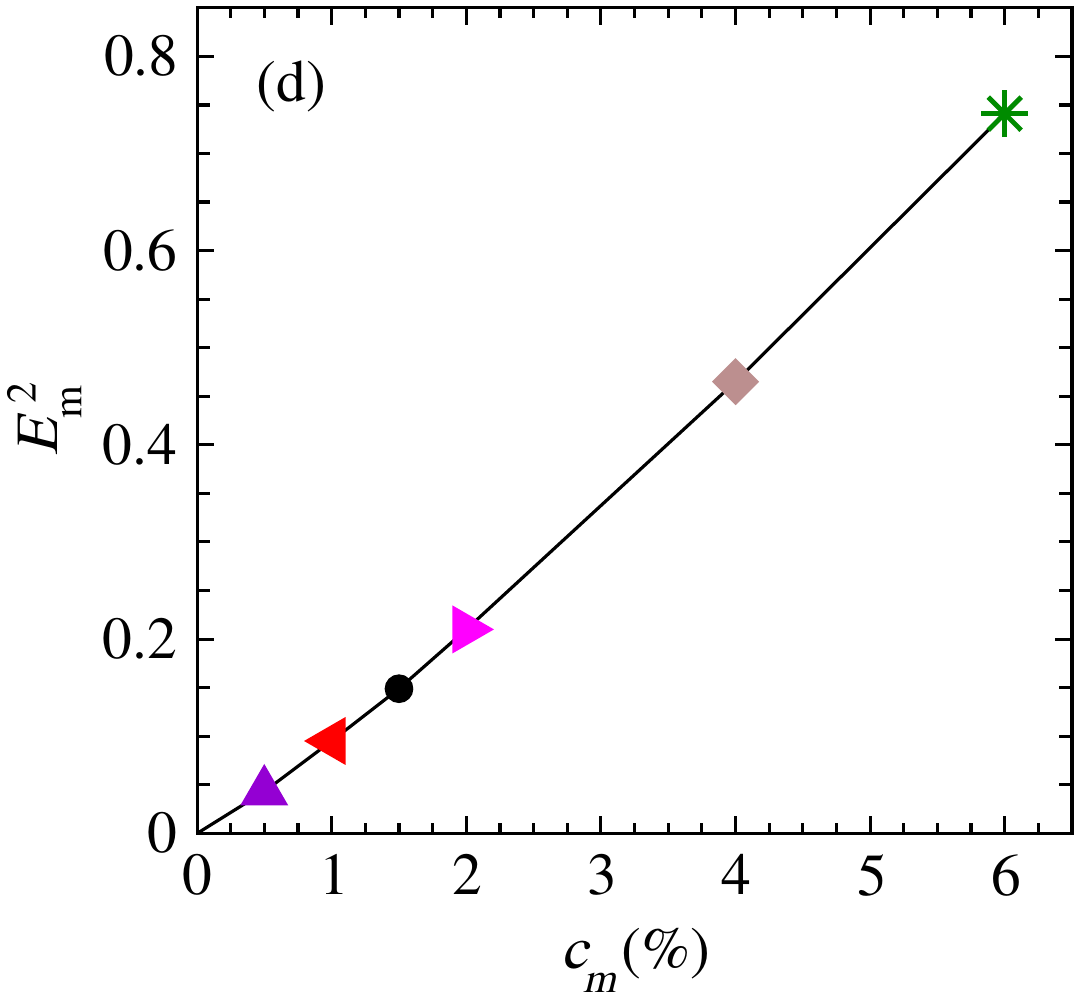}

\end{center}

\caption{ \label{Fig_DOS_Sig_d1A1B1}
(color online)
Electronic density of states and conductivity computed from TB1 in BLG: 
with (a.1) (a.2)  A$_2$ vacant atoms, 
and with (b.1) (b.2) B$_2$ vacant atoms:
(a.1) (b.1) total DOS $n(E)$ (dashed lines TDOS without vacancy);
(a.2) (b.2) microscopic conductivity $\sigma_m(E)$.
$c$ is the concentration of vacancies with respect to the total number of atom in BLG.
(c) TDOS of MLG with A vacant atoms and (d) the corresponding $E_m^2$ value versus the concentration $c_m$ of vacancies with respect to the total number of atom in MLG. 
As explained in the supplementary material \cite{supplementaryMat} (Sect. 2 and 4), theses plots do not include the midgap states at $E_D = 0$. 
Spectrum is symmetric with respect to Dirac energy $E_D=0$.
}
\end{figure}

The total density of states for both layers (TDOS), $n(E)$,   
are shown figures \ref{Fig_DOS_Sig_d1A1B1}(a.1) and \ref{Fig_DOS_Sig_d1A1B1}(b.1) respectively for A and B vacant atoms in layer 2.
As explained in the following and in supplementary material \cite{supplementaryMat} (Sect. 2),
each missing orbital in the A$_2$ sublattice (resp.  B$_2$ sublattice) of the top layer 2 produces one midgap states at Dirac energy $E_D = 0$ that spreads on \{A$_1$, B$_2$\} sublattices 
(resp.  \{A$_2$, B$_1$\}). This is  similar to the case of a monolayer of graphene where vacancies in sublattice A (resp. B) produce midgap states at Dirac energies $E_D = 0$ that are located in sublattice B (resp. A) \cite{Peres06,Pereira08a}. 
In figures \ref{Fig_DOS_Sig_d1A1B1}(a.1), \ref{Fig_DOS_Sig_d1A1B1}(b.1) and \ref{Fig_DOS_Sig_d1A1B1}(c) the midgap states at $E_D = 0$ are not included in plotted DOSs 
and in the calculation of the conductivity
(supplementary material \cite{supplementaryMat} Sect. 2 
and 4). 
Vacancies on the A$_2$ sublattice do not produce a gap in the TDOS, whereas B$_2$ vacancies induce a quasi-gap clearly seen around $E_D=0$. 
Its width, $\Delta E_{g} = 2 E_b$, increases when $c$ increases and saturates at a value $2t_1$.
In B$_2$ vacancies case, unphysical small oscillations appeared in the total DOS and local DOSs. Those oscillations are numerical artifacts related to the termination of the continuous fraction expansion of the Green function used in the recursion method (see supplementary material \cite{supplementaryMat} Sect. 2, 3, and Ref. \cite{methodeRecursion}). The presence of these unphysical oscillations in the case of B$_2$ vacancies whereas there is no oscillations in the cases of A$_1$ vacancies, confirms the emergence of a gap by B$_2$ vacancies.

The conductivity $\sigma_m(E)$, is shown figures \ref{Fig_DOS_Sig_d1A1B1}(a.2) and \ref{Fig_DOS_Sig_d1A1B1}(b.2) for A$_2$ vacant atoms and B$_2$ vacant atoms, respectively.
In both cases, the conductivity at  large energies $|E| \gg 1$\,eV is inversely proportional to  the concentration $c$ of vacancies. This is expected from the  Boltzmann theory if the vacancies are seen only as scattering centers which give a finite lifetime to the eigenstates of the perfect Bernal bilayer. For smaller energies, corresponding to usual $E_F$ values, the variation of the conductivity with the concentration $c$ of vacancies is not consistent with  Boltzmann theory. 
Indeed, 
with vacancies on A$_2$ sublattice, for small $E$ values, $\sigma(E)$ increases strongly when $c$ increases. 
With vacancies on B$_2$ sublattice, 
for energies above the quasi-gap, i.e. $E > E_b$,  if  $c < c_l \simeq 1.5$\,\%, $\sigma(E)$ decreases when $c$ increasses (as expected in Boltzmann theory); 
whereas for $c > c_l$, $\sigma(E)$ increases when $c$ increases.

All these spectacular results show that the effect of selective functionalization is not just to induce scattering for the states of the perfect bilayer. This is also confirmed by analyzing the selective functionalization of a sublattice of the MLG. As shown in figure \ref{Fig_DOS_Sig_d1A1B1}(c),  it leads to the creation of a quasi-gap which width increases with concentration of adatoms. Let us recall that for a monolayer and bilayer with vacancies that are randomly distributed on the two sublattices A and B   (Refs. \cite{Ducastelle13,Roche13,Trambly13,Missaoui17} and Refs. there in) the low energy  DOS presents  a  peak which is reminiscent of the midgap states  but  has a finite width.

\subsection{Results with Hamiltonian including hopping beyond nearest neighbor (TB2)}
\begin{figure}
\begin{center}

\includegraphics[width=4.1cm]{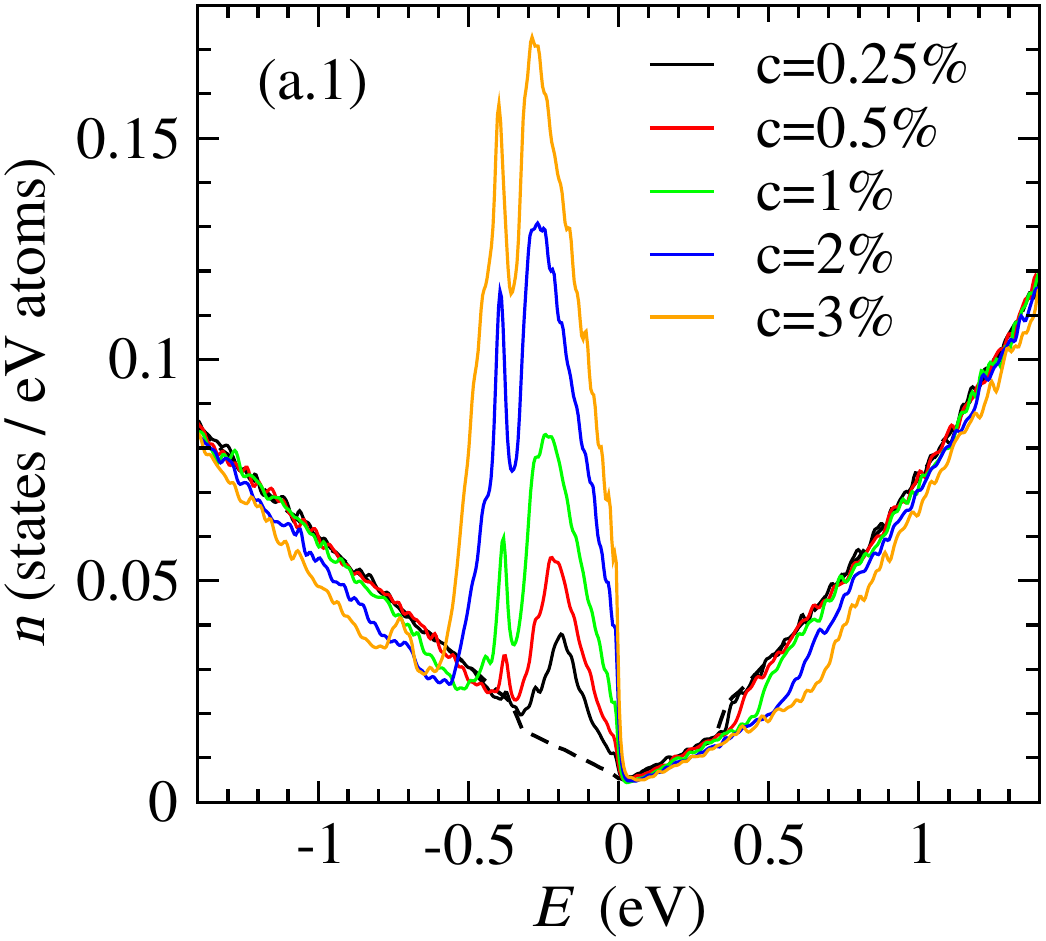} 
~ \includegraphics[width=4.1cm]{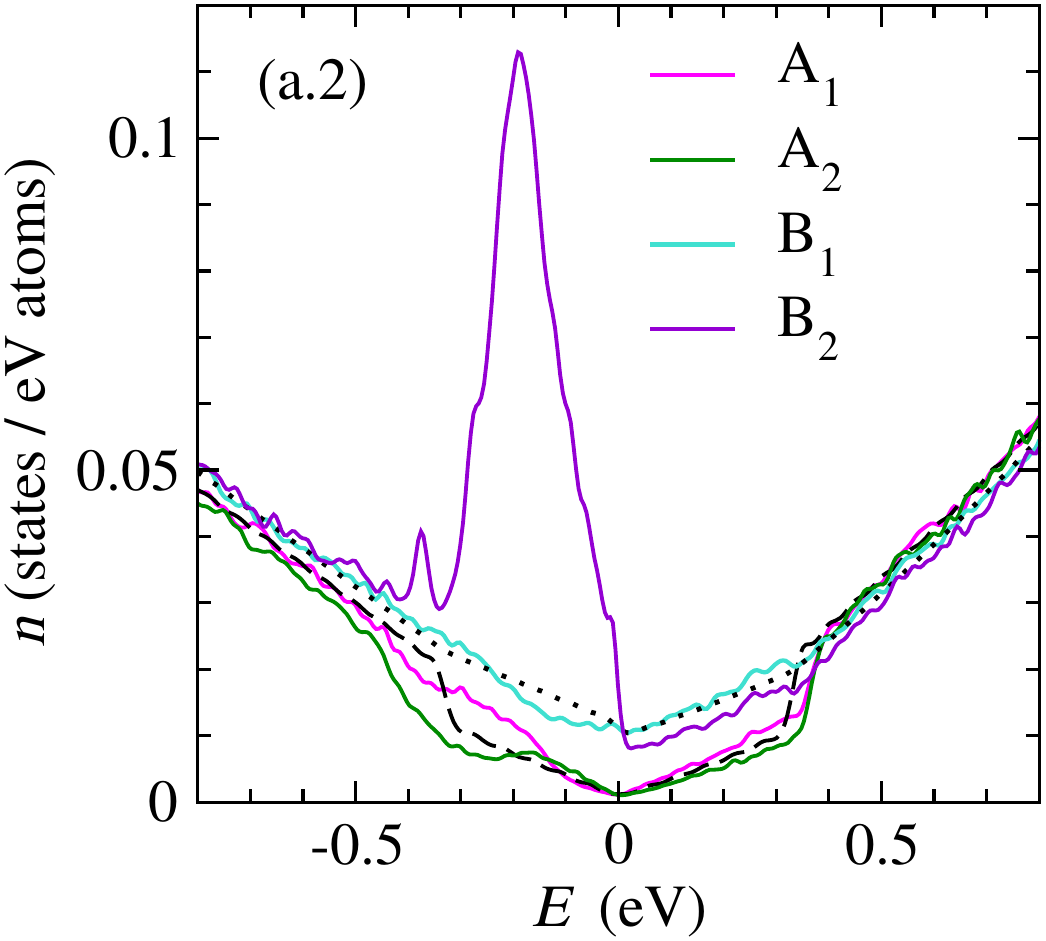} 

\vspace{.1cm}
\includegraphics[width=7.6cm]{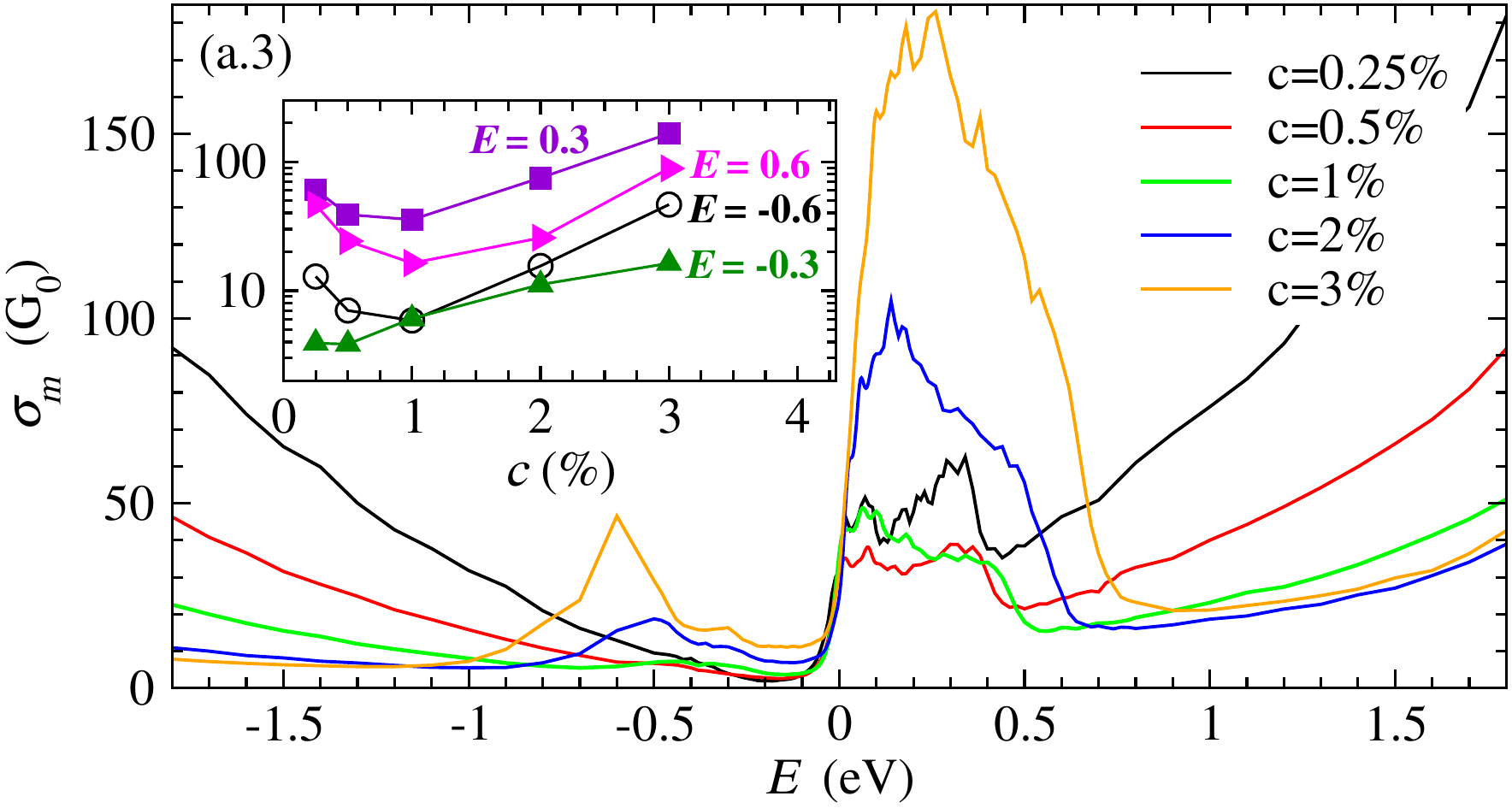} 

\vspace{.1cm}
\includegraphics[width=4.1cm]{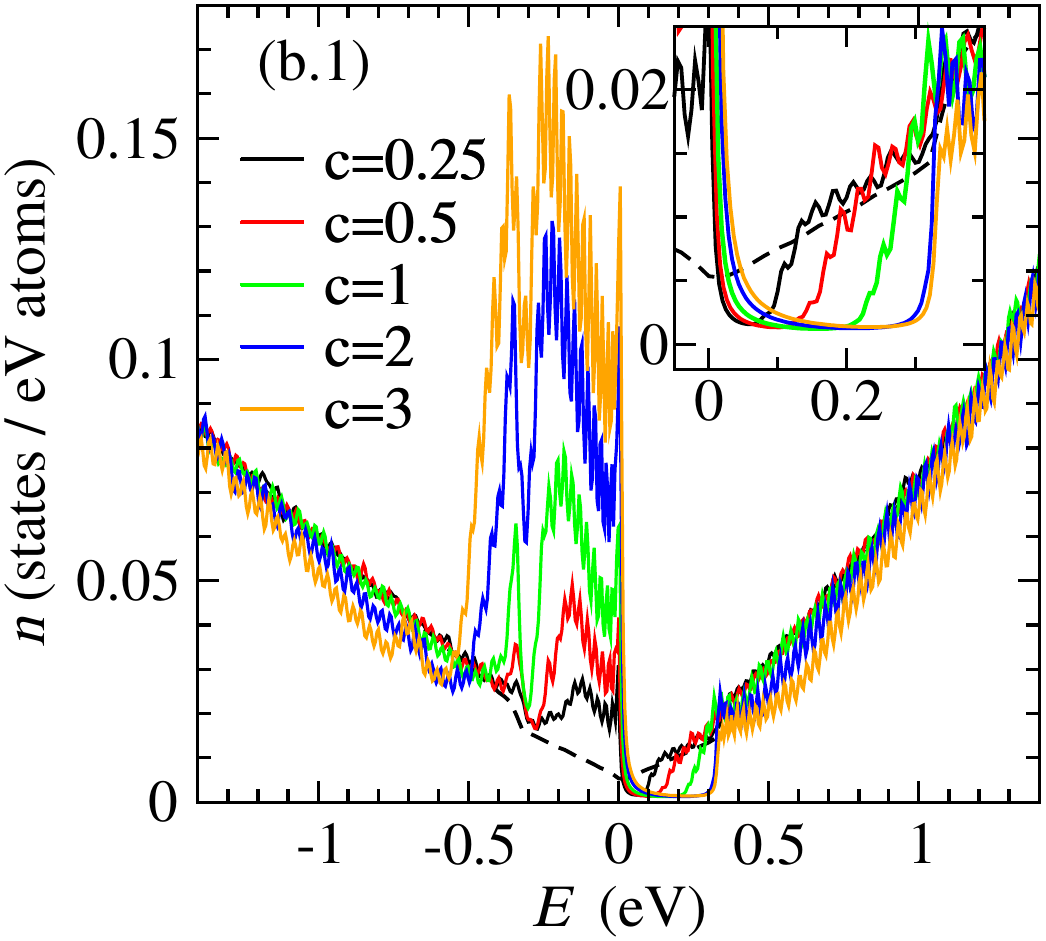} 
~ \includegraphics[width=4.1cm]{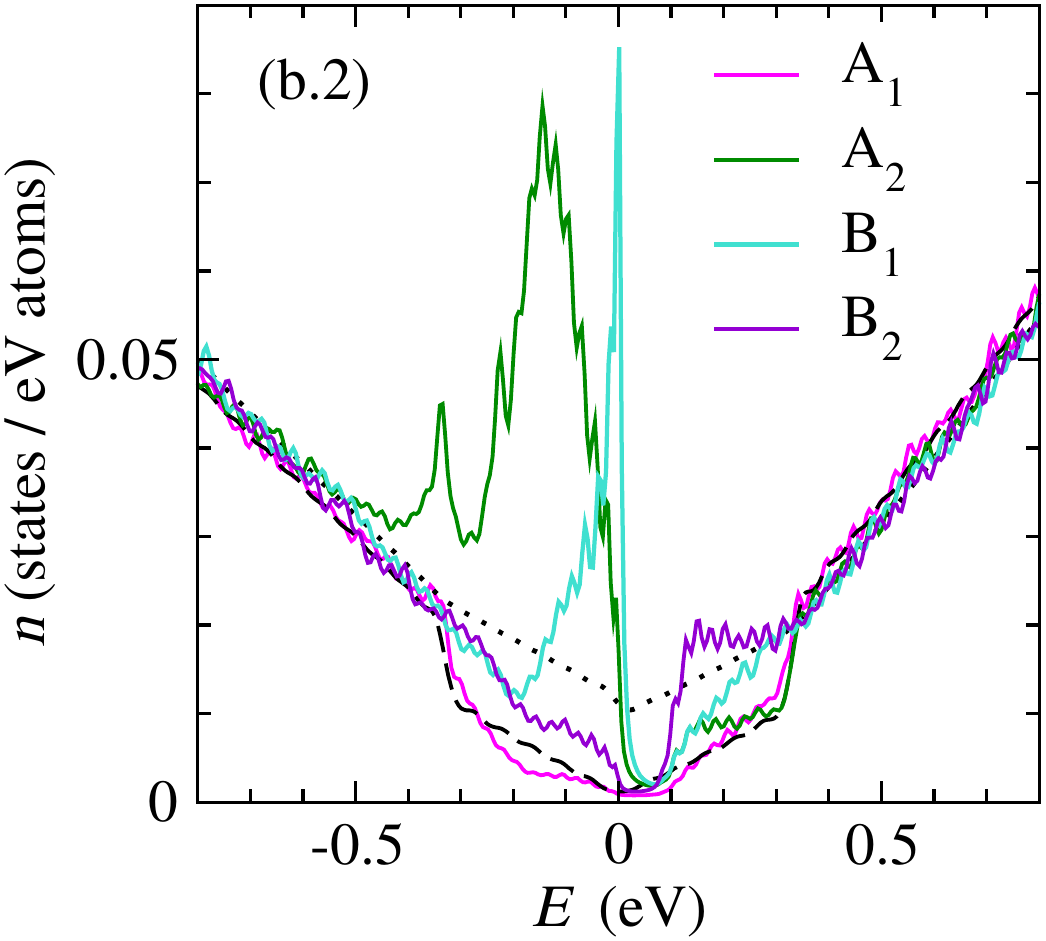} 

\vspace{.1cm}
\includegraphics[width=7.6cm]{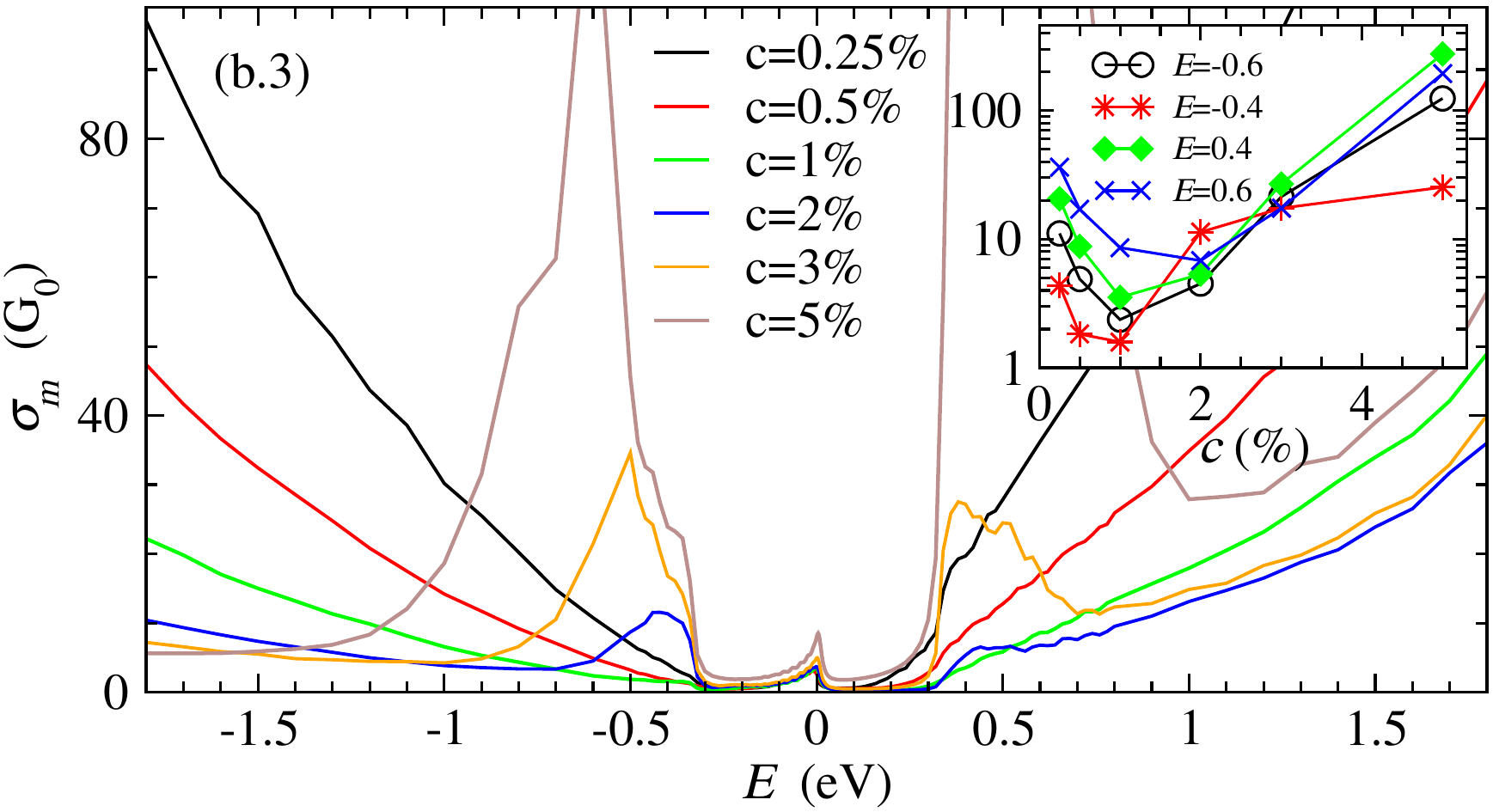} 

\end{center}

\caption{ \label{Fig_DOS_Sig_TB2}
(color online)
Electronic properties computed from TB2  (including hopping terms beyond nearest neighbor) in BLG with (a) A$_2$  vacant atoms and (b) B$_2$ vacant atoms:
(a.1) (b.1) total DOS (dashed line is the total DOS without vacancy, and insert shows the DOS arround $E=0$),
(a.2) (b.2) average local DOS on A$_1$, B$_1$, A$_2$, B$_2$ atoms for $c = 0.25$\% 
(dashed line and dot line are LDOS on A and B atom without vacancy),
(a.3) (b.3) microscopic conductivity $\sigma_m(E)$
[The insert shows the conductivity $\sigma_m$ versus $c$ for several $E$ values (eV)].
$c$ is the concentration of vacancies with respect to the total number of atom in BLG. 
}
\end{figure}

Now we present results calculated using TB2 model, including hopping beyond nearest neighbors, in place of TB1 model (supplementary material \cite{supplementaryMat} (Sect. 1)). 
The TDOS, $n(E)$, the average LDOS, $n_i(E)$ with $i=$A$_1$, A$_2$, B$_1$, B$_2$,  
and the conductivity, $\sigma_m(E)$, are shown in figures \ref{Fig_DOS_Sig_TB2} for A$_2$ vacant atoms and B$_2$ vacant atoms.
In both cases the midgap states, produced by missing orbitals are displaced to negative energy by the effect of the hopping beyond nearest neighbors (TB2) as in MLG \cite{Pereira08a,Trambly11,Trambly14} and BLG with vacancies randomly distributed \cite{Missaoui17}. In addition these states appear in an energy window of a fraction of an eV that depends on the concentration of functionalized sites. 
It is interesting to note that the peak of vacancy states is split into a double peak when we increase the concentration of vacancies. That splitting indicates a coupling between vacancy states that are all located on the same sublattice.
The symmetry of the electronic properties with respect to $E_D=0$ of TB1 model is  broken; but, qualitatively, the anomalous conductivity found in the case of TB1 model is still found with TB2 model. The main difference between TB1 and TB2 is in the energy window where  the midgap states appear.

With A$_2$ vacant atoms,
the average LDOS (figure \ref{Fig_DOS_Sig_TB2}(a.2)) shows that midgap states is located on $B_2$ orbitals of the same layer, as expected from the uncompensated theorem with TB1. 
For $c \ge c_l \simeq 1$\%, $\sigma_M(E)$ increases strongly when $c$ increasses (figure \ref{Fig_DOS_Sig_TB2}(a.3)). 
This increase is maximum (several order of magnitude) for energies close to $-0.6$ and $0.3$ eV, and it is smaller for energies corresponding to the midgap peak. 

With B$_2$ vacant atoms,
the average LDOS (figure \ref{Fig_DOS_Sig_TB2}(b.2)) shows that midgap states is located on $A_2$ orbitals (layer 2) and on $B_1$ orbitals (layer 1), as expected from a bipartite Hamiltonian analysis with TB1 \cite{Missaoui_these}. 
The quasi-gap is found for $0 < E \le E_b$, instead of $-E_b < E \le E_b$ with TB1, since for negative energies midgap states are present in the case of TB2.
$\sigma_m$ is very small for $E$ corresponding to midgap states, and theses energies correspond at a mobility quasi-gap 
(figure \ref{Fig_DOS_Sig_TB2}(b.3)).  
As a result, similarly to TB1 model, a mobility quasi-gap is found for $-E_b < E \le E_b$ with TB2 too. 
Moreover for  $c \le c_l \simeq 1$--$1.5$\% and for $E > E_b$, $\sigma(E)$ increases strongly when $c$ increases (as with TB1).

\section{Discussion: Interpretation of the results by bipartite lattice.}
We analyze now the origin  for  the formation of the gap in the MLG with selective functionalization on sublattice A (or B) and then show how it leads to the  properties of the BLG. 
Quite generally an eigenstate with energy $E$ of the MLG  with or without vacancies,  can be writen as 
$| \varphi \rangle= | \varphi_A \rangle +| \varphi_B \rangle$, 
with states  $| \varphi_A \rangle $ ($ | \varphi_B \rangle $) belonging to the  sublattice  $A$ ($B$). 
It is easy to  show that
$\varphi_A$ and $\varphi_B$ are eigenstates of the effective Hamiltonian $\hat{\tilde{H}} = \hat{H}^2 $ with eigenvalue $\tilde{E} = E^2$. $\hat{\tilde{H}}$ acts only within the sublattices $A$ and $B$ and does not couple them. 

For example for the perfect MLG,  $\hat{\tilde{H}}$ is the Hamiltonian of a triangular lattice of A atoms (B atoms),
\begin{equation}
\hat{\tilde{H}}_{A}  = \sum_i \tilde{\epsilon}_{A} c^{\dag}_{Ai}c_{Ai} 
+ \sum_{\langle i,j \rangle} \tilde{t}_{0}  c_{Ai}^{\dag}c_{Aj} + h.c. ,
\label{eq_TBT}
\end{equation}
where $c_{Ai}^{\dag}$ and $c_{Ai}$ creates and annihilates respectively a state of an electron $A_i$, and
with $\tilde{\epsilon}_{A} = 3 t_0^2$ and $\tilde{t}_{0} = t_0^2$.
The middle of $\hat{{H}}$ band ($E=0$) corresponds to the lowest energy of $\hat{\tilde{H}}$ band ($\tilde{E}= E^2=0$). 


The effect of vacancies on the DOS away from zero energy  can be understood by considering the effective Hamiltonian of the sublattice $A$ that contains the vacancies. This Hamiltonian has the  form  given in equation (\ref {eq_TBT}) but with functionalized sites that are simply deleted. Without vacancies the coordination $\eta$ of each atoms of A sublattice is 6. With a small concentration $c_m$ of vacancies in A sublattice, the average coordination is $\eta \simeq 6(1-c_m[\%]/100)$. The center of the A band is fixed by on-site energies, $\tilde{\epsilon}_{A}$, and it is not affected by vacancies; but the width of the band will decrease  when $\eta$ decreases (i.e. when $c_m$ increases). 
As expected from this simple tight-binding argument, the minimum values, $\tilde{E}_{m}= E^2_{m}$, of the spectrum of $\hat{\tilde{H}}$, found numerically (figure \ref{Fig_DOS_Sig_d1A1B1}(d)),  is almost proportional to the average coordination number $\eta$ (average number of A--A (B--B)  nearest neighbors of A (B) sublattice of the bipartite lattice). 
Consequently the average A DOS, $\tilde{n}_{A}$, has a gap induced by vacancies for $0 \le E^2 \le E^2_{m}$. 
This means that
DOS in the A and B sublattices of MLG also presents a gap for $-E_{m} \le E \le E_{m}$ (figure \ref{Fig_DOS_Sig_d1A1B1}(c)) 
As is well known each vacancy in sublattice $A$  also induces a zero energy midgap states in sublattice $B$. Note that similar results are obtained on a square lattice which is also a bipartite lattice (supplementary material \cite{supplementaryMat}, Sect. 3).



Let us consider now  the case of the bilayer with vacancies on the A$_{2}$ sublattice. In this case the midgap states of the top layer 2 are located only on sublattice B$_{2}$ and are not coupled to the lower layer 1. Therefore layer 1 is just coupled to a semi-conductor (top layer 2) with a gap in the energy range  $ -E_{m} \le E \le E_{m}$. The results shown above mean that $t_1 $ is sufficiently small that the mixing between states of  layer 1 and 2  is small. Therefore layer 1 has essentially the electronic structure of an isolated MLG without defects. This explains why the TDOS is similar to that of a graphene layer.  In addition when the vacancies concentration increases, $E_{m}$ increases and the decoupling between the two layers is more efficient. Therefore at a given energy  the lifetime of states in the lower layer 1 increases and  the conductivity increases when concentration increases. Transport in the bilayer at these energies $ -E_{m} \le E \le E_{m}$ is mainly through the lower layer 1.

The case of vacancies on the B$_{2}$ sublattice is slightly more complex.  Again at energies $E$ such that $ -E_{m} \le E \le E_{m}$ the mixing between states of layer 1 and states in the continuum of layer 2 is small. However  in that case the midgap states of layer 2 are located on sublattice A$_{2}$ and are coupled to the sublattice A$_{1}$ of the  lower layer 1. The effect of the interlayer coupling alone is to couple midgap states of  A$_{2}$ with  specific linear combinations of states of A$_{1}$ and to produce bonding and anti-bonding states at energies $ t_1$ and $-t_1$.  We consider now the case where the concentration of adatoms is sufficient to have $ E_{m} \ge t_1 $. Therefore at energies $E$ such that $ -t_{1} \le E \le t_{1}$ these  specific states in sublattice A$_{1}$ appear as decoupled from the other states of layer 1. They act thus as vacancies in the MLG (layer 1)  and this produces a quasi-gap with midgap states in sublattice B of layer 1. For that reason, a quasi-gap exists in both layers  in the energy range $ -t_1 \le E \le t_1 $ and it is seen in the TDOS. Similarly to the previous case, increasing the concentration can also increase the conductivity for energies $E$ such that  $ t_1 \le |E|  \le E_{m}$. 

These analyses of the effect of selective functionalization are confirmed by detailed studies of the bipartite Hamiltonian of BLG \cite{Missaoui_these}.

\section{Conclusion}
We have analyzed the density of states and the conductivity of graphene Bernal bilayer (BLG) when the upper layer is functionalized by adatoms. Since there are two inequivalent sublattices A and B, that correspond to carbon atoms that are more or less coupled to the lower layer, we study the effect of a selective functionalization of sublattices A or B. As we show this selective functionalization  leads to the creation of a gap when sublattice B of the upper layer is randomly  functionalized with a concentration  of adatoms $c\geq 10^{-2}$. This gap is a fraction of one eV for the DOS and of at least $0.5$\,eV  for the mobility. This phenomenon is intimately related to the bipartite structure of the graphene lattice and the maximum width of the gap is of the order of the interlayer coupling energy. Other functionalizations of sites are possible if both layers can be functionalized. In this case also we find that electronic structure and transport properties can be deeply modified by a selective functionalization \cite{Missaoui_these}. 
We believe that the phenomenon due to selective functionalization could be observed in carefully prepared graphene bilayers or even in other 2D materials.

\section*{Acknowledgments}
The authors wish to thank C. Berger, W. A. de Heer, L. Magaud, P. Mallet and J.-Y. Veuillen
for fruitful discussions.
The numerical calculations have been performed 
at  Institut N\'eel, Grenoble,
and at the Centre de Calculs (CDC),
Universit\'e de Cergy-Pontoise.
We thank Y. Costes, CDC, for computing assistance. 
This work was supported by the Tunisian French Cooperation Project (Grant No. CMCU 15G1306)
and the project ANR-15-CE24-0017.

\newpage

\begin{center}
\large{{\bf Supplementary Material}  }
\end{center}

\renewcommand{\thefigure}{S\arabic{figure}}
\setcounter{figure}{0}

In this supplementary material, we first (section\,1) present the tight-binding models for Bernal bilayer graphene (BLG): TB1 that includes only the first neighbor hoppings, 
and a more realistic model, TB2, that includes hopping beyond first neighbors. 
In section 2, we discuss midgap states at energy $E=E_D=0$ with the density of states (DOS) of BLG calculated with TB1. 
Section 3 presents the electronic structure of a bipartite square lattice with vacancies in a sublattice. 
The method to compute Kubo-Greenwood conductivity is described section 4.

\section*{1. Tight-binding Hamiltonian Models}
\label{Sec_TMmodel}

In this part, we present in details the 
tight-binding (TB) schemes.
Bilayer graphene (BLG) consists in four carbon atoms in its unit cell, two carbons A$_1$, B$_1$ in layer 1 and A$_2$, B$_2$ in layer 2 where A$_2$ lies on the top of A$_1$.
Only $p_z$ orbitals are taken into account since we are interested in what happens at the Fermi level. 
The Hamiltonian has the form :
\begin{equation}
\hat{H} = \sum_{\langle i,j \rangle} t_{ij} \left( c_i^{\dag}c_j + c_j^{\dag}c_i \right),
\label{Eq_hamiltonian_SupMat}
\end{equation}
where $c_i^{\dag}$ and $c_i$ create and annihilate respectively an electron on the $p_z$ orbital located at $\vec r_i$ ,
$\langle i,j\rangle$ is  the sum on index $i$ and $j$ with $i\ne j$, 
and  $t_{ij}$ is the hopping matrix element
between two $p_z$ orbitals located at $\vec r_i$ and $\vec r_j$.
We consider two model Hamiltonians. 

\begin{figure}
\begin{center}
\includegraphics[width=6.5cm]{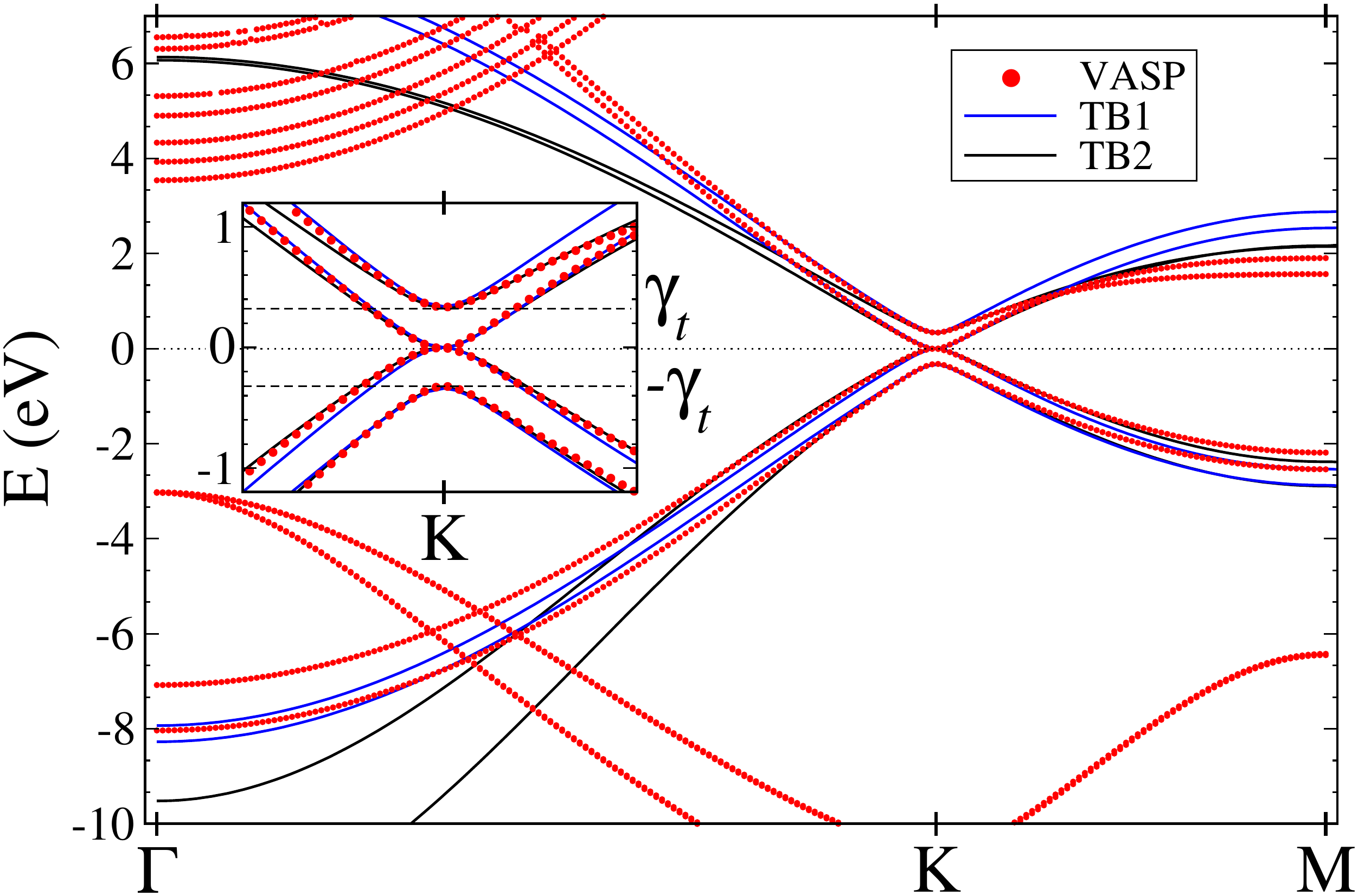}
\caption{ \label{Fig_DOS_bnds_AB}
Bands dispersion in bilayer graphene (BLG) computed with the two model hamiltonians: 
(blue line) model with only first-neighbor hopping (TB1),
(black line) model with hopping above first-neighbor hopping (TB2).
Red points are ab-initio results from VASP method (from \cite{Trambly12}).
}
\end{center}
\end{figure}

The first model (TB1 model) is the simplest TB model with first-neighbor hopping only. 
Each atom has 3 first neighbors in the same plane with hopping between p$_z$ orbitals $t_0 = 2.7$\,eV. 
The inter-layer hopping between  p$_z$ orbitals between A$_1$ and first neighbor A$_2$  is $t_1 = 0.34$\,eV
to reproduce similar band dispersion as in ab-initio calculations (figure \ref{Fig_DOS_bnds_AB}).
Inter-layer hopping splits two bands of MLG Dirac cone in two parabolic bands separated by 
$2 \gamma_t$ (figure \ref{Fig_DOS_bnds_AB}). With the TB1 model, $\gamma_t = t_1 = 0.34$\,eV, 
the spectrum is symmetric with respect to Dirac energy $E_D$, $E_D = 0$.

The second model (TB2 model) is more realistic where hopping terms are not restricted to nearest neighbor hopping. 
We have used this model Hamiltonian in our previous works \cite{Trambly10,Trambly12,Trambly16,Missaoui17} to study electronic structure 
of the rotated graphene bilayers. It reproduces the {\it ab initio} calculations 
of the electronic states for energies within $\pm 1$\,eV of $E_{\rm D}$ (figure \ref{Fig_DOS_bnds_AB}).
The hopping terms are computed from Slater-Koster parameters,
\begin{eqnarray}
t_{ij} 
~=~   n_c^2 V_{pp\sigma}(r_{ij}) ~+~ (1 - n_c^2) V_{pp\pi}(r_{ij}),
\end{eqnarray}
where $n_c$ is the direction cosine of 
$\vec r_{ij} = \vec r_j - \vec r_i$ along $(Oz)$ axis 
and $r_{ij}$ is the distance between the orbitals,
\begin{eqnarray}
n_c ~=~ \frac{z_{ij}}{r_{ij}} ~~~{\rm and}~~r_{ij} = ||\vec r_{ij}|| .
\end{eqnarray}
$z_{ij}$ is the coordinate of $\vec r_{ij}$ along $(Oz)$. $z_{ij}$  is either equal to zero or 
to a constant because the two graphene layers have been kept flat in our model.
We use the same dependence on distance of the Slater-Koster parameters:
\begin{eqnarray}
V_{pp\pi}(r_{ij})    &=&-\gamma_0 \,{\rm e}^{q_{\pi}    \left(1-\frac{r_{ij}}{a} \right)} , \label{eq:tb0}\\
V_{pp\sigma}(r_{ij}) &=& \gamma_1 \,{\rm e}^{q_{\sigma} \left(1-\frac{r_{ij}}{a_1} \right)} .
\label{eq:tb1}
\end{eqnarray}
where $a$ is the nearest neighbor distance within a layer, 
$a=1.418$\,{\rm \AA}, and $a_1$ is the
interlayer distance, $a_1=3.349$\,{\rm \AA}. 
First neighbor interaction in
a plane is taken equal to the commonly used value, $\gamma_0 = 2.7$\,eV \cite{Castro09_RevModPhys}.
Second neighbor interaction $\gamma_0'$ 
in a plane is set \cite{Castro09_RevModPhys}  to
$0.1\times\gamma_0$; 
that fixes the value of the ratio $q_{\pi}/{a}$ in equation (\ref{eq:tb0}).
The inter-layer hopping between two $p_z$ orbitals in $\pi$ configuration 
is $\gamma_1$. $\gamma_1$ is fixed to obtain a good fit with ab-initio 
calculation around Dirac energy in AA stacking  \cite{Trambly12}
and AB bernal stacking  and then to get $\gamma_t=0.34$\,eV 
(figure \ref{Fig_DOS_bnds_AB}) 
which results in $\gamma_1=0.48$\,eV.
We choose the same coefficient of the exponential decay 
for $V_{pp\pi}$ and $V_{pp\sigma}$,
\begin{eqnarray}
\frac{q_{\sigma}}{a_1} ~=~ \frac{q_\pi}{a} 
~=~ \frac{{\ln} \left(\gamma_0'/\gamma_0 \right)}{a - a_0}
~=~ 22.18{\rm \,nm}^{-1},
\end{eqnarray}
with $a_0 = 2.456$\,{\rm \AA} the distance between second
neighbors in a plane. 

We consider that resonant adsorbates --simple atoms or molecules such as H, OH, CH$_3$--  create a covalent bond with some carbon atoms of the BLG. To simulate this covalent bond, we assume that the $p_{z}$ orbital of the carbon, that is just below the adsorbate, is removed. In our calculations, the mono-vacancies are distributed at random on one sublattice, i.e. on one type of atom in one layer, with a finite concentration $c$.

\begin{figure}
\begin{center}
\includegraphics[width=4.1cm]{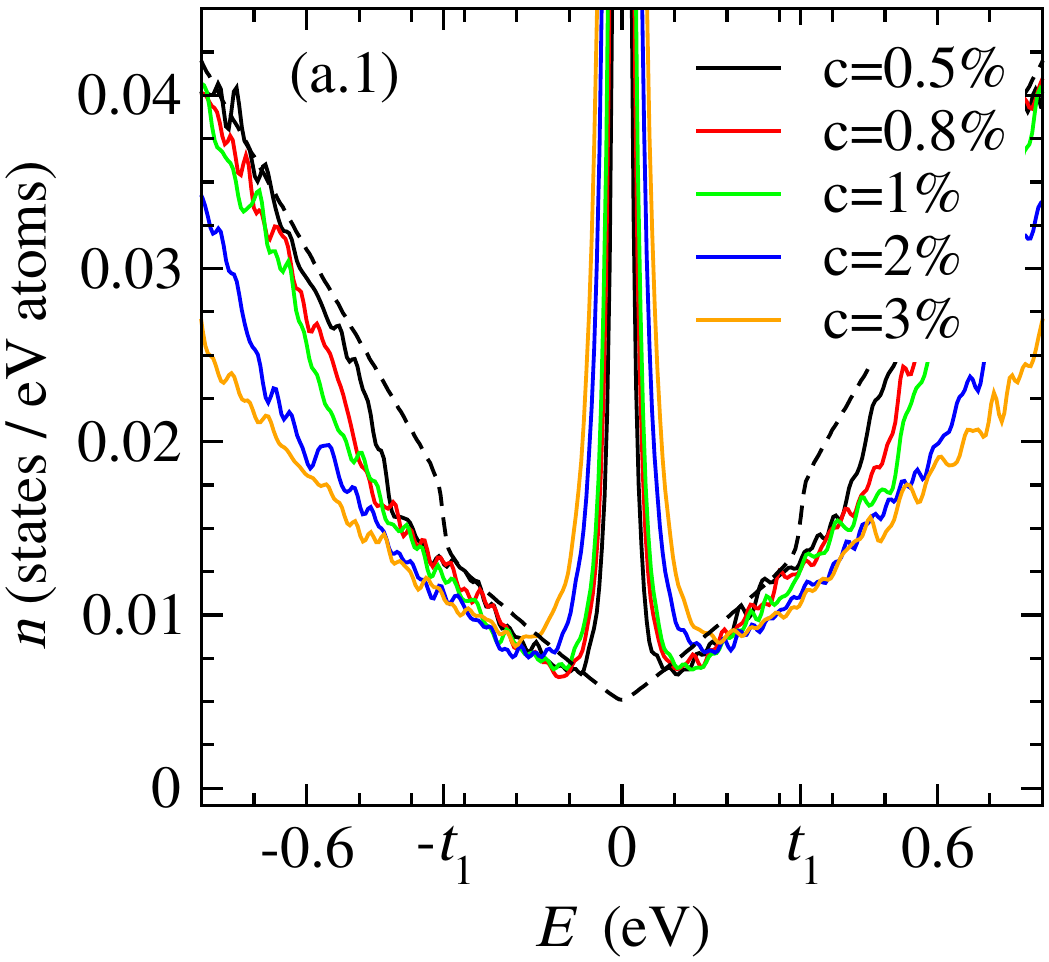} 
~ \includegraphics[width=4.1cm]{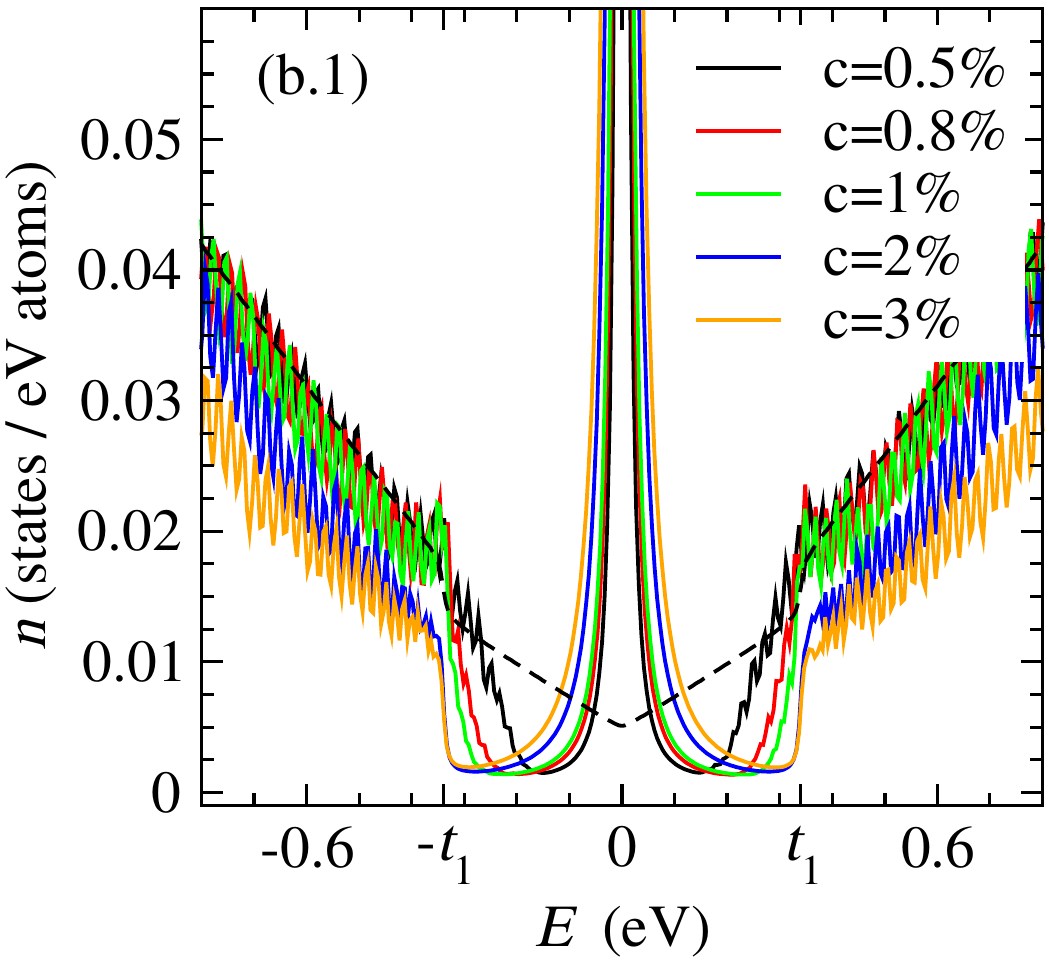}~ 

\vspace{.1cm}
\includegraphics[width=4.1cm]{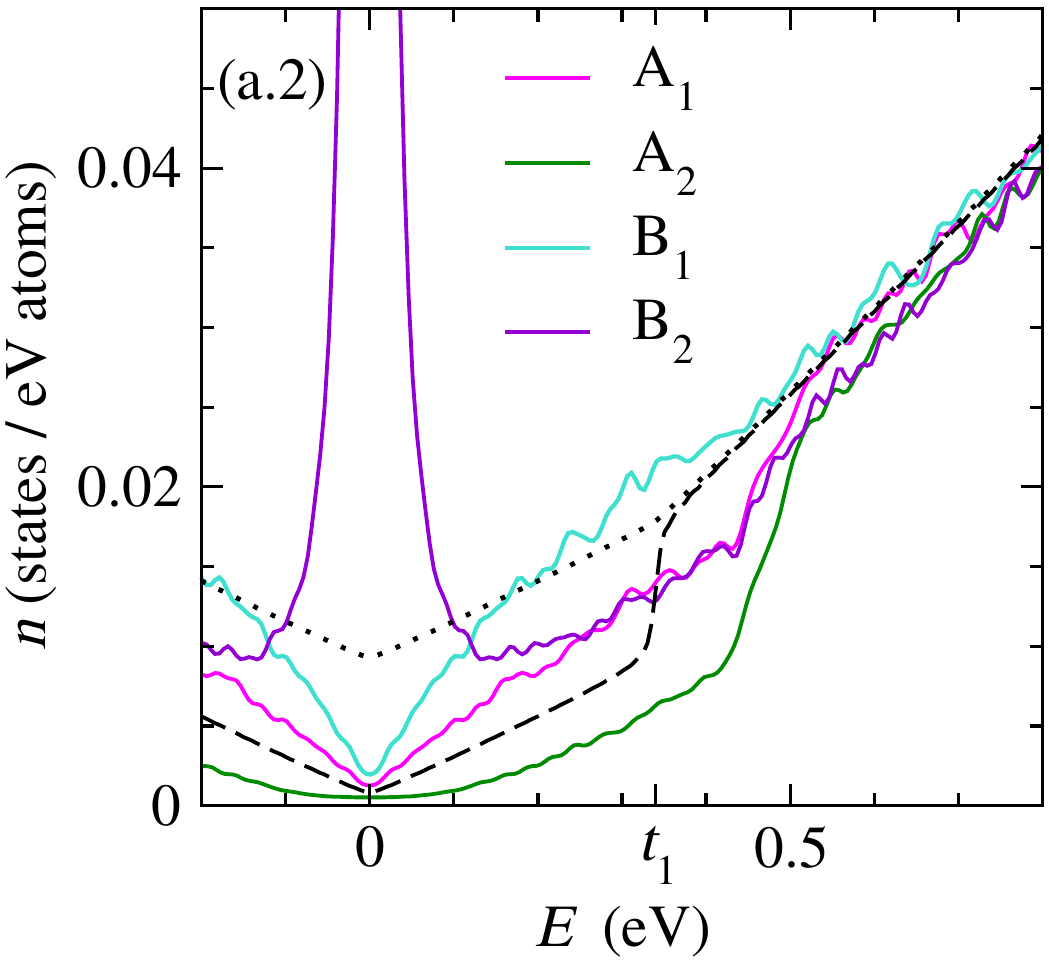} 
~ \includegraphics[width=4.1cm]{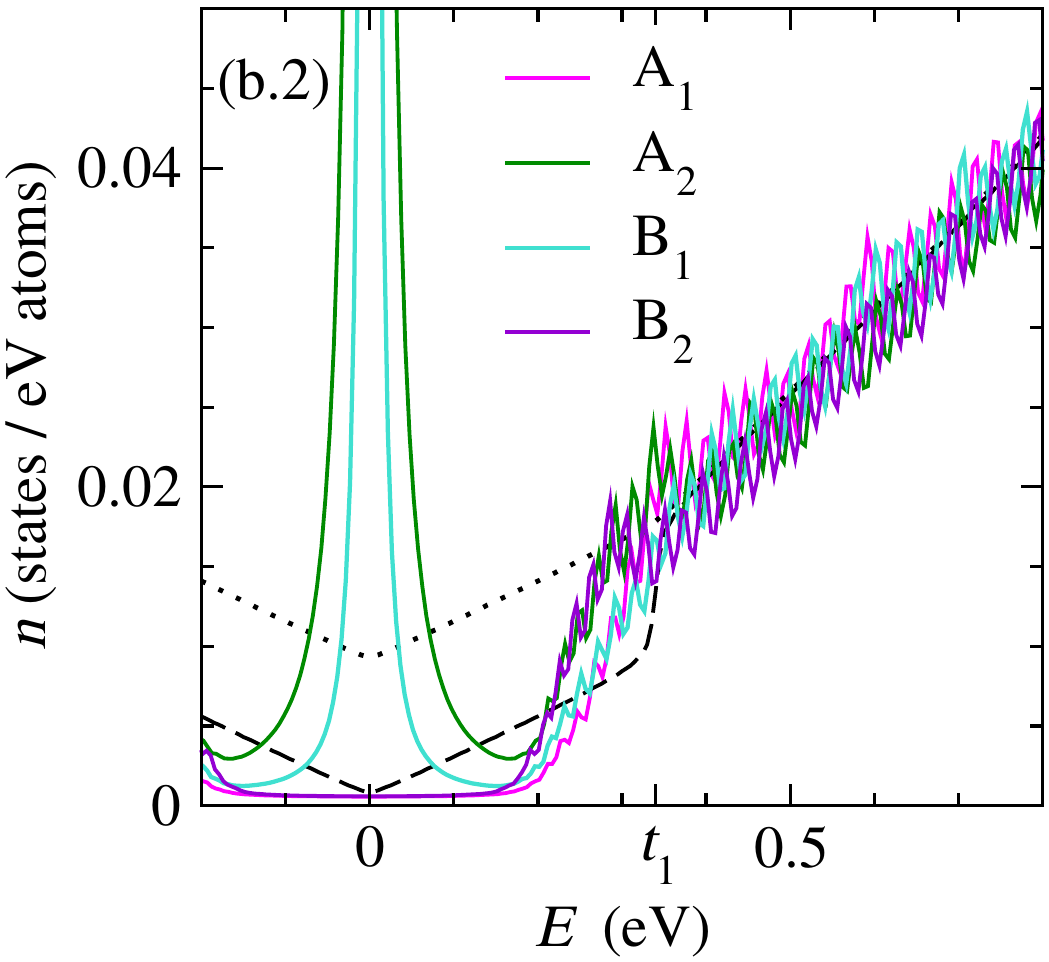} 

\vspace{.1cm}

\caption{ \label{Fig_DOS_d1A1_B1}
Electronic structure of BLG with TB1 (nearest neighbor hopping):
In BLG with A$_2$ vacant atoms:
(a.1) total DOS, $n'$  (dashed line is TDOS without vacancy),
(a.2) average local DOS $n_i'$ on $i=$ A$_1$, B$_1$, A$_2$, B$_2$ atoms with $c=0.5$\% (dashed line and dot line are LDOS on A and B atom without vacancy).
In BLG with B$_2$ vacant atoms:
(b.1) total DOS $n'$ (dashed line is TDOS without vacancy),
(b.2) average local DOS $n_i'$ on $i=$ A$_1$, B$_1$, A$_2$, B$_2$ atoms $c=0.5$\% 
(dashed line and dot line are LDOS on A and B atom without vacancy),
$c$ is the concentration of vacancies with respect to the total number of atom in BLG. 
The spectrum is symmetric with respect to Dirac energy $E_D=0$.
DOS are calculated from (\ref{eq_nepsilon}) with $\epsilon = 5$\,meV. 
}

\vskip 0.5cm 

\includegraphics[width=4.1cm]{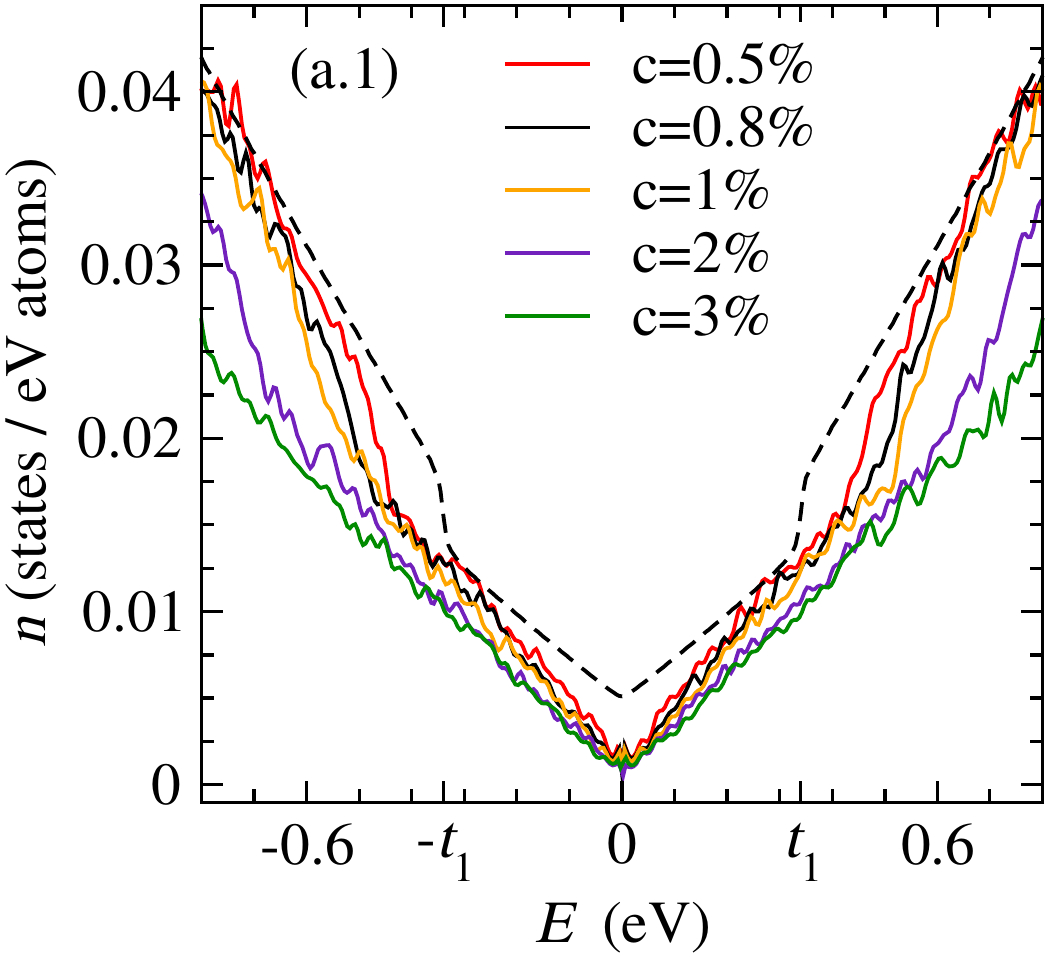} 
~ \includegraphics[width=4.1cm]{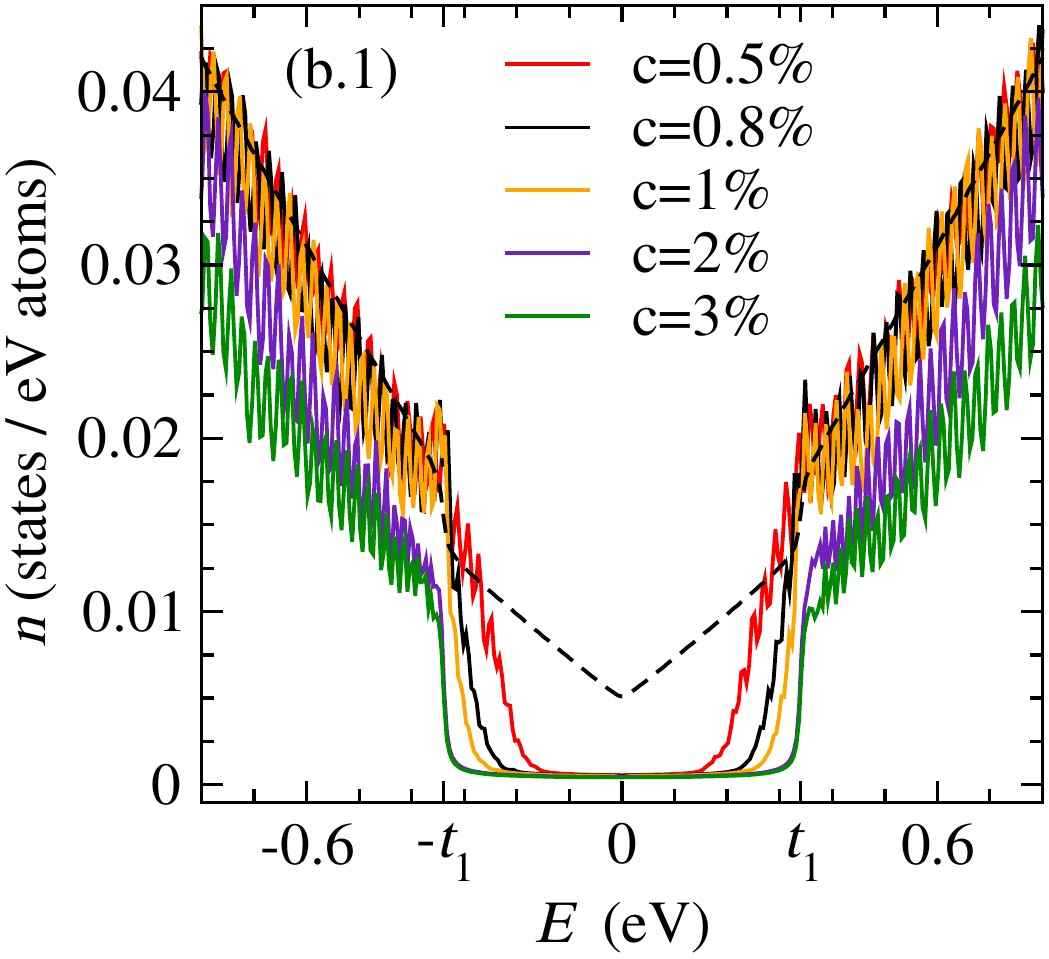}~ 

\vspace{.1cm}
\includegraphics[width=4.1cm]{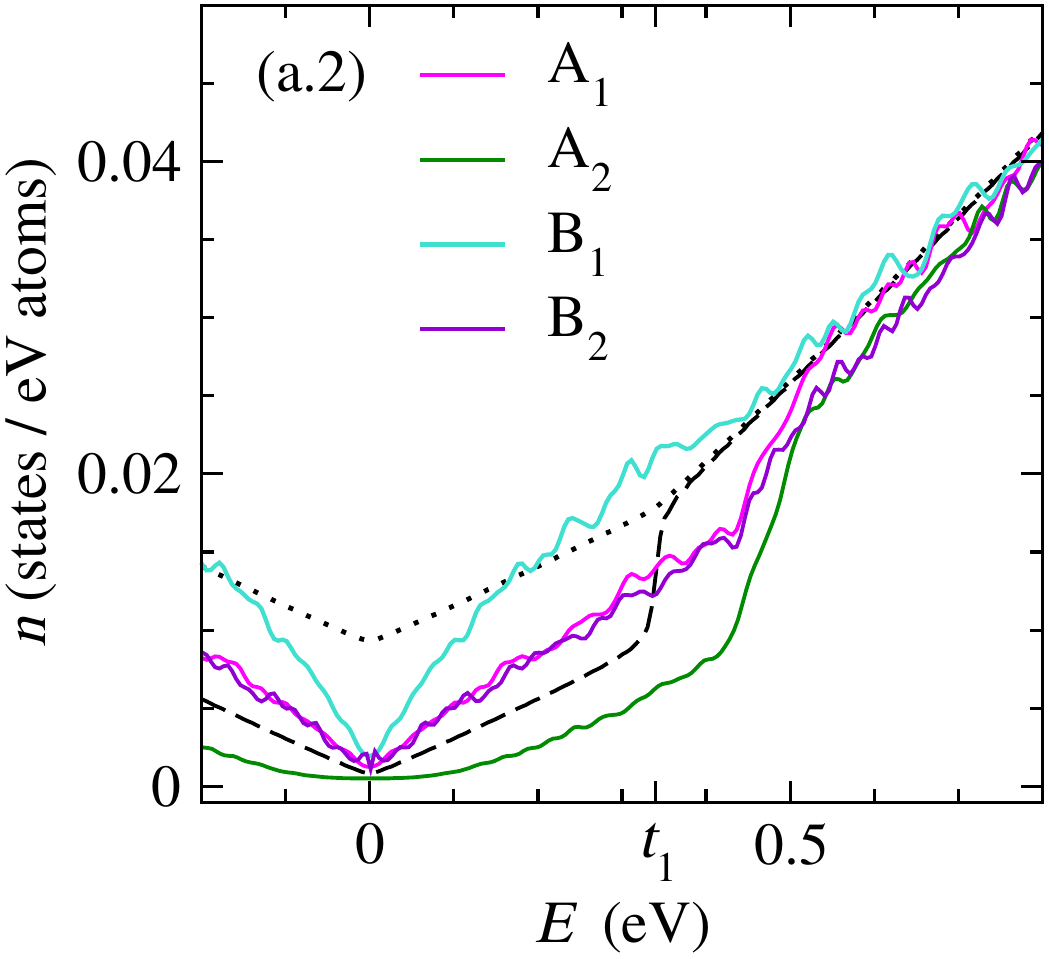} 
~ \includegraphics[width=4.1cm]{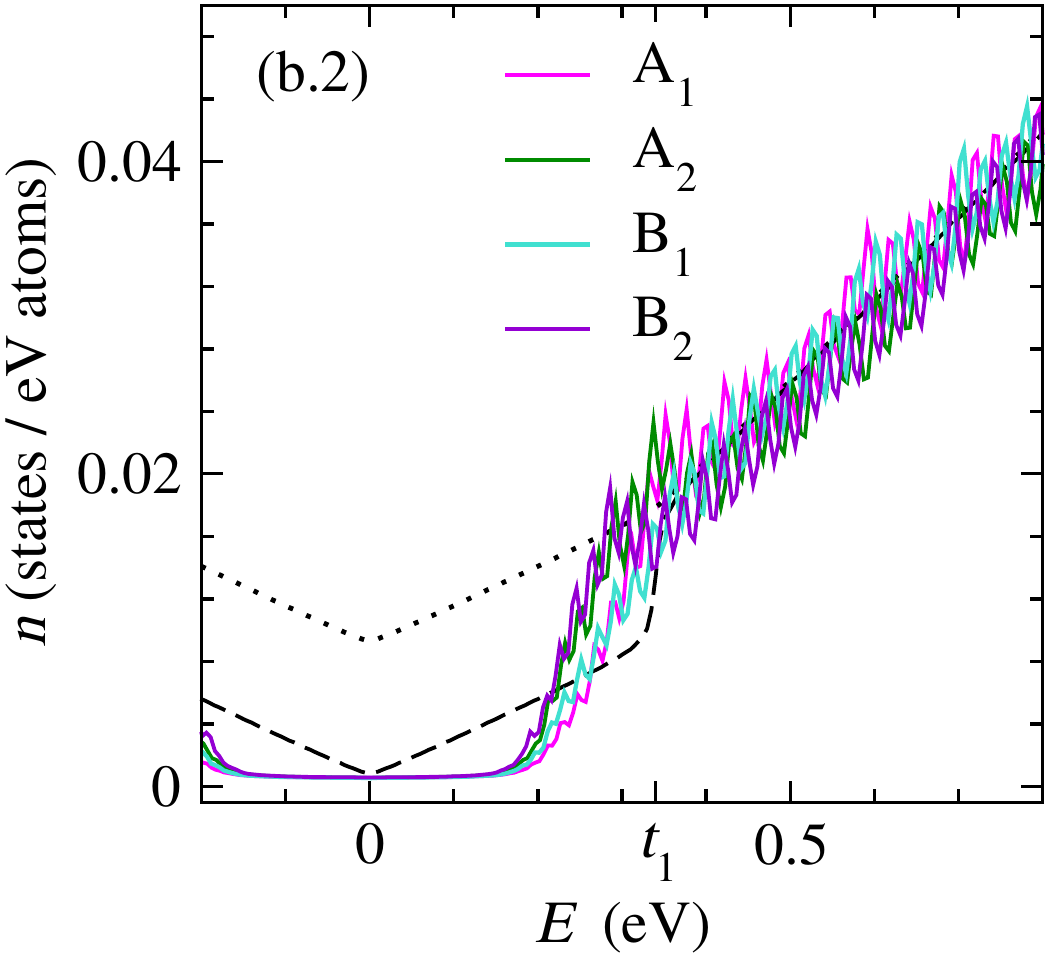} 

\caption{ \label{Fig_DOS_d1A1B1_mLor}
Same DOS and LDOS than figure \ref{Fig_DOS_d1A1_B1}, but without drawing the midgap states at $E=0$.
$n$ (without midgap) is computed from $n'$ (with midgap states) by equations (\ref{Eq_DOS_avec_sans_Lor}) and (\ref{Eq_DOS_avec_sans_Lori}). 
}
\end{center}
\end{figure}

\section*{2. Midgap states in case of first neighbor hopping model (TB1) with vacancies on a sublattice only}
\label{subLorentzienne}

\begin{figure}
\begin{center}
\includegraphics[width=6cm]{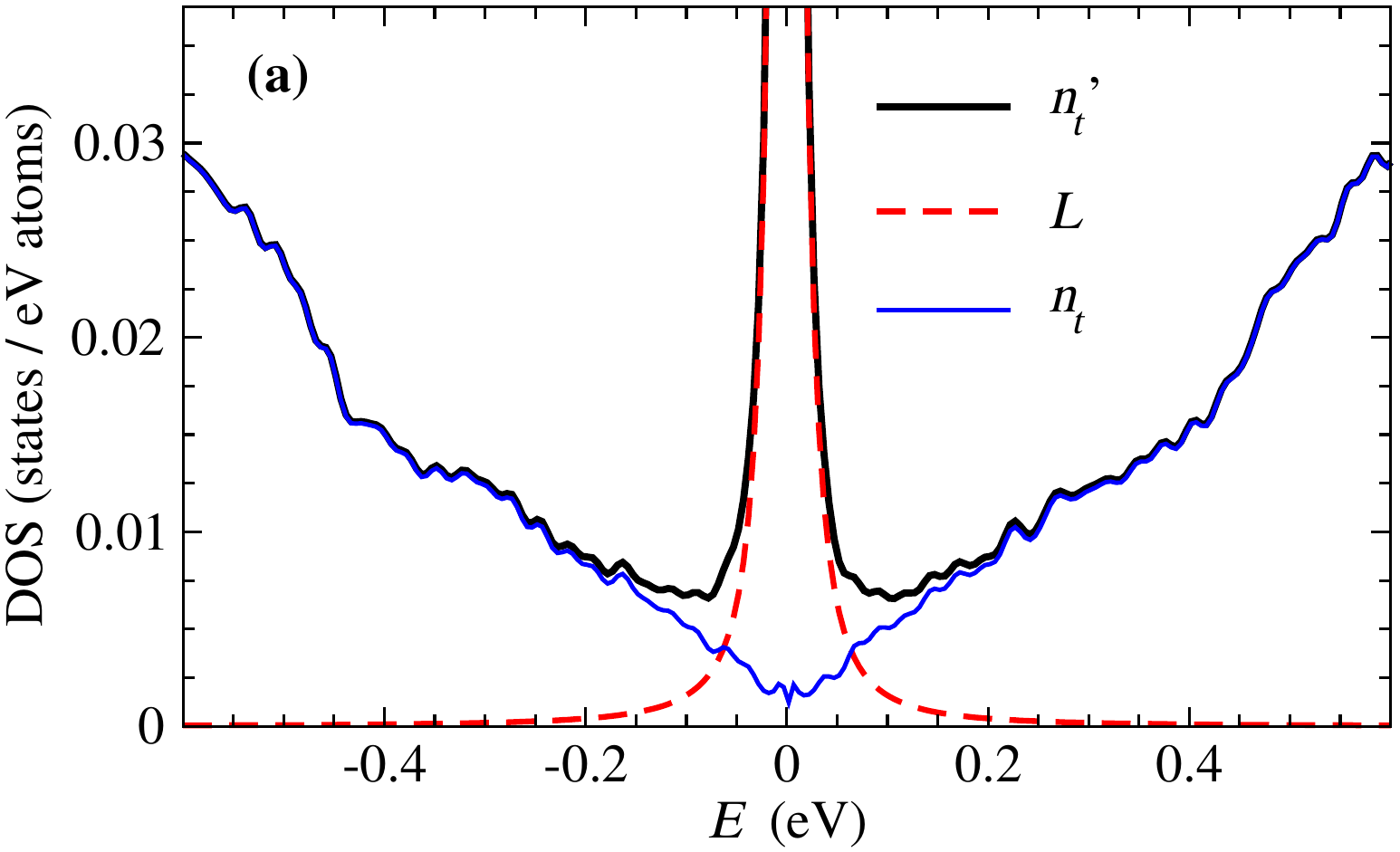} 

\vspace{.1cm}
\includegraphics[width=6cm]{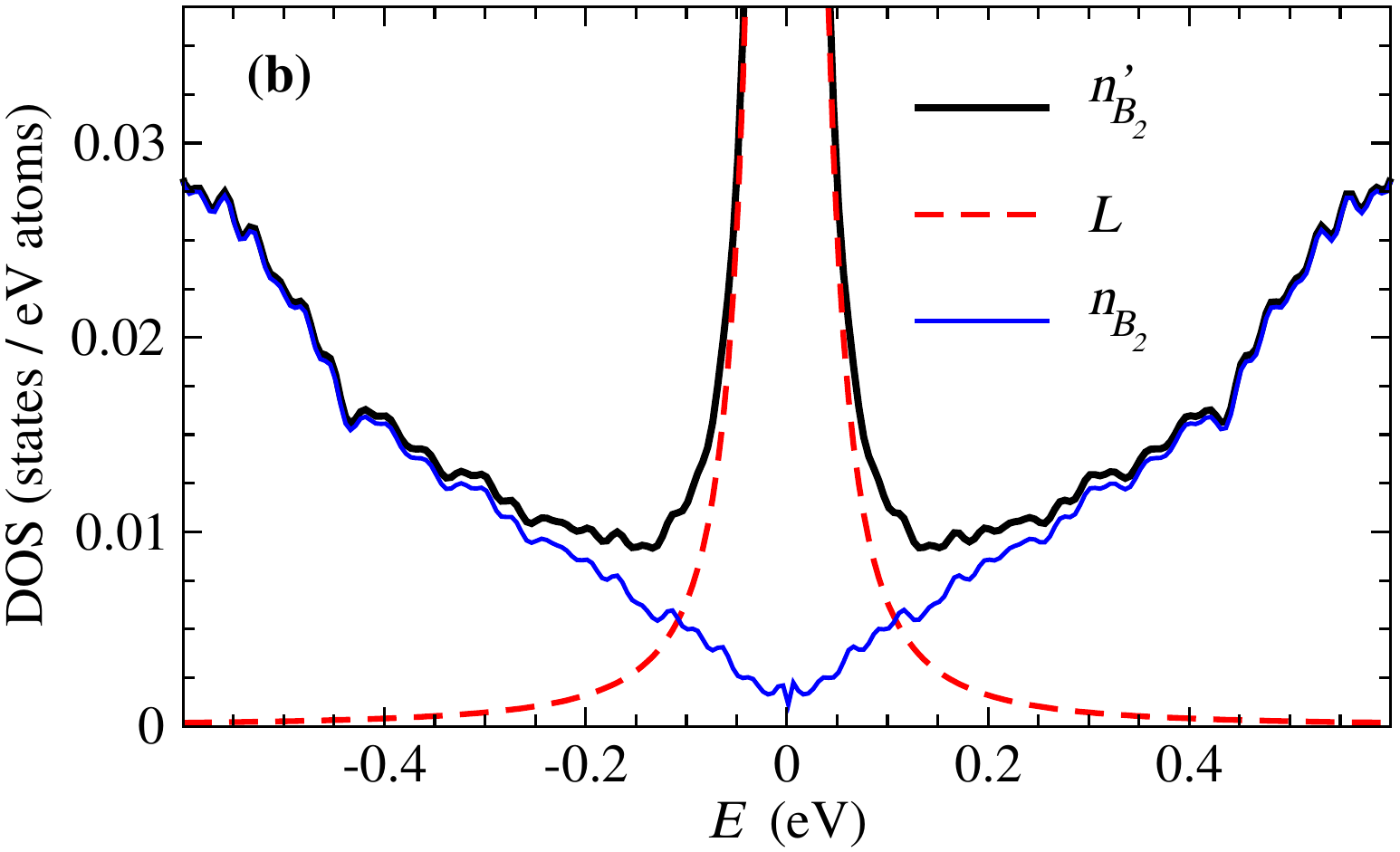} 

\caption{Electronic density of states in BLG  with A$_2$ vacant atoms ($c=0.5$\%): 
(a) Calculated total DOS $n'(E)$,  $n'(E)=n(E) + x L(E)$, $x=0.01$;
(b) Calculated average local DOS on B$_2$ atoms $n_{B_2}'(E)$, $n_{B_2}'(E)=n_{B_2}(E) + x_{2} L(E)$, $x_2 = 0.04016$,
where $L(E)$ is the Lorentzian of the midgap states at $E=0$ that are located only on B$_2$ sublattice, and
$n(E)$ and $n_{B_2}(E)$ are, respectively, the exact total DOS and the exact average local DOS on B$_2$ sublattice at all energies except at $E=0$. 
\label{Fig_DOS_LaA1_avec_sans_Lor}
}
\end{center}
\end{figure}

\begin{figure}
\begin{center}
\includegraphics[width=6cm]{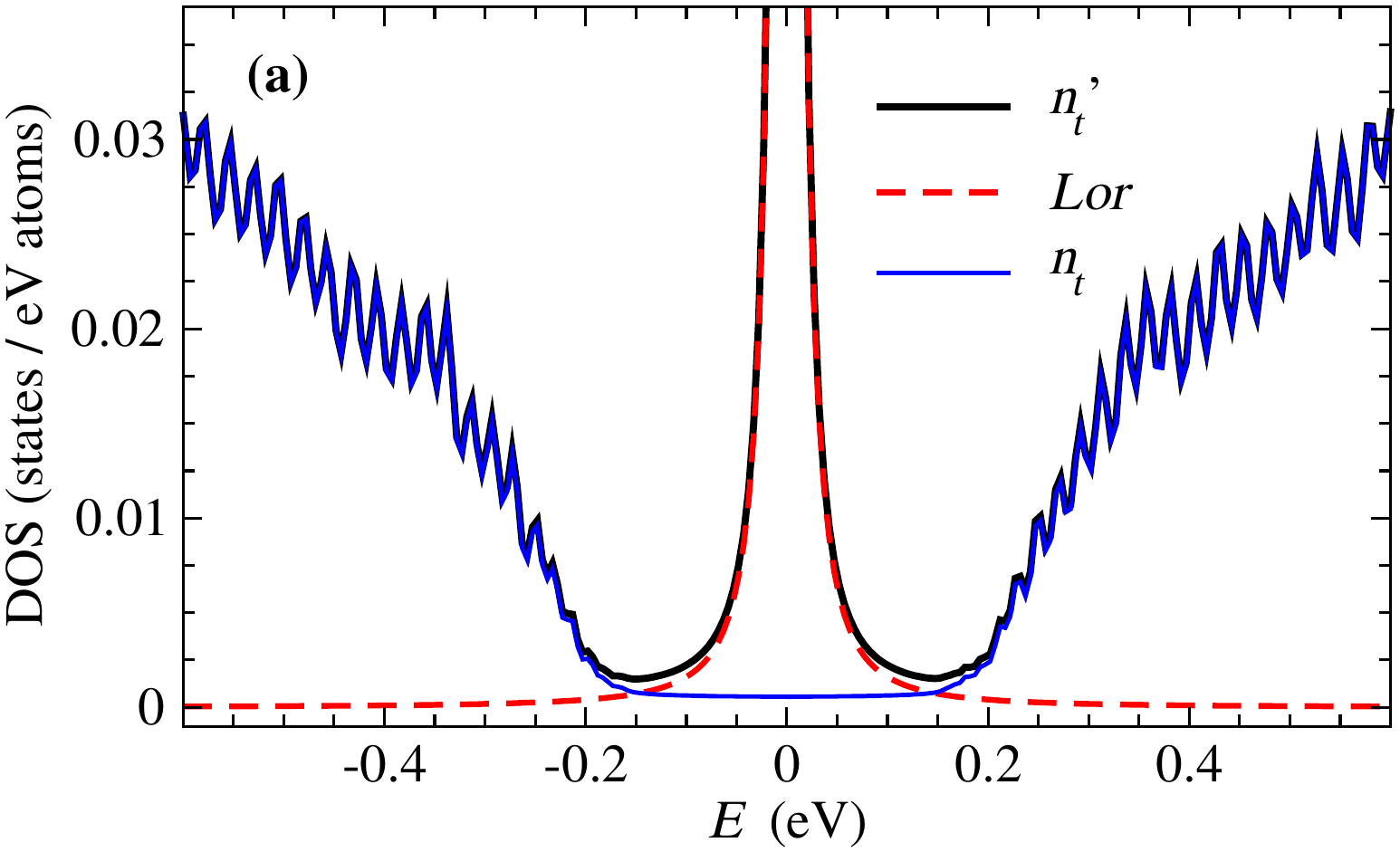} 

\vspace{.1cm}
\includegraphics[width=6cm]{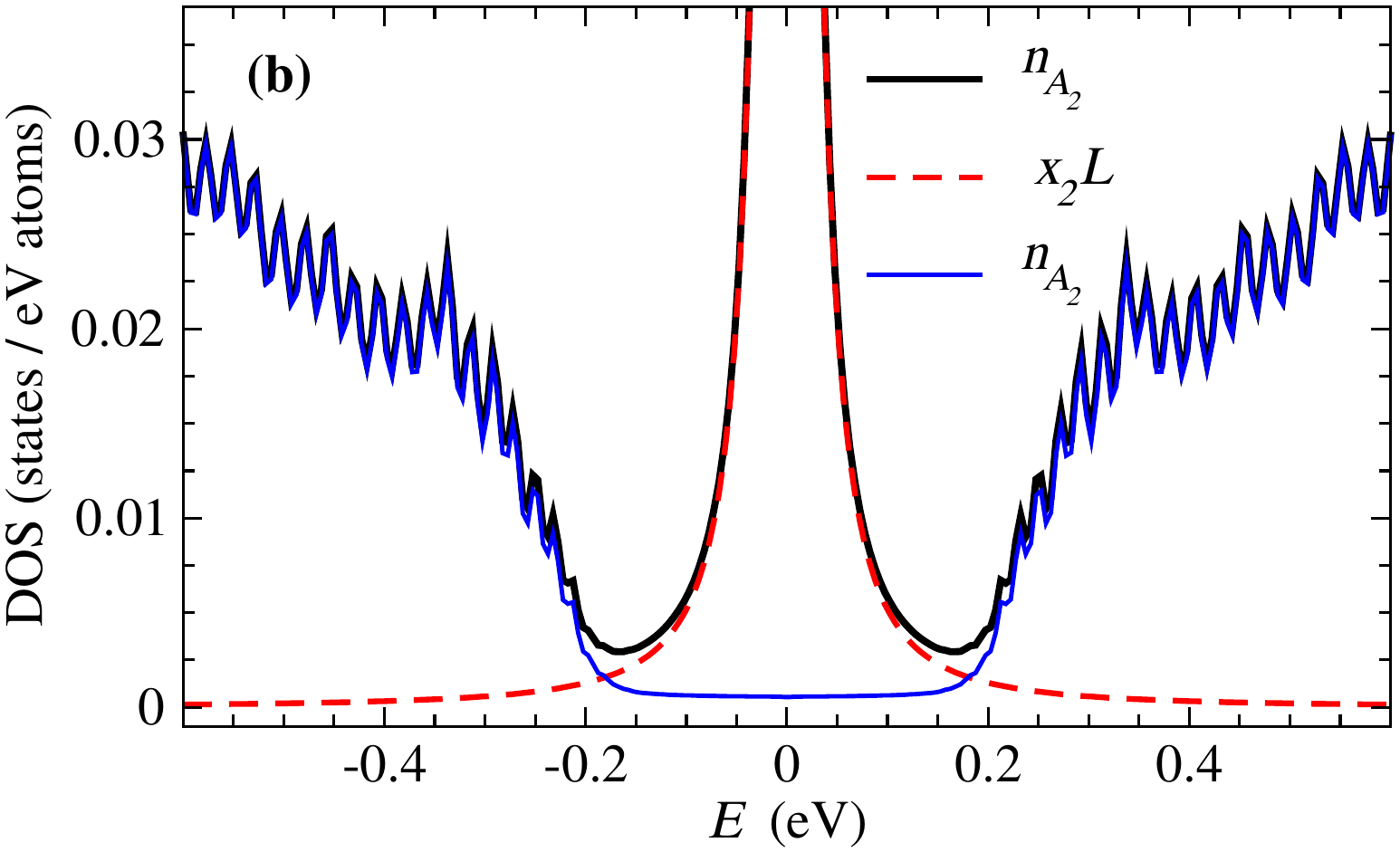} 

\vspace{.1cm}
\includegraphics[width=6cm]{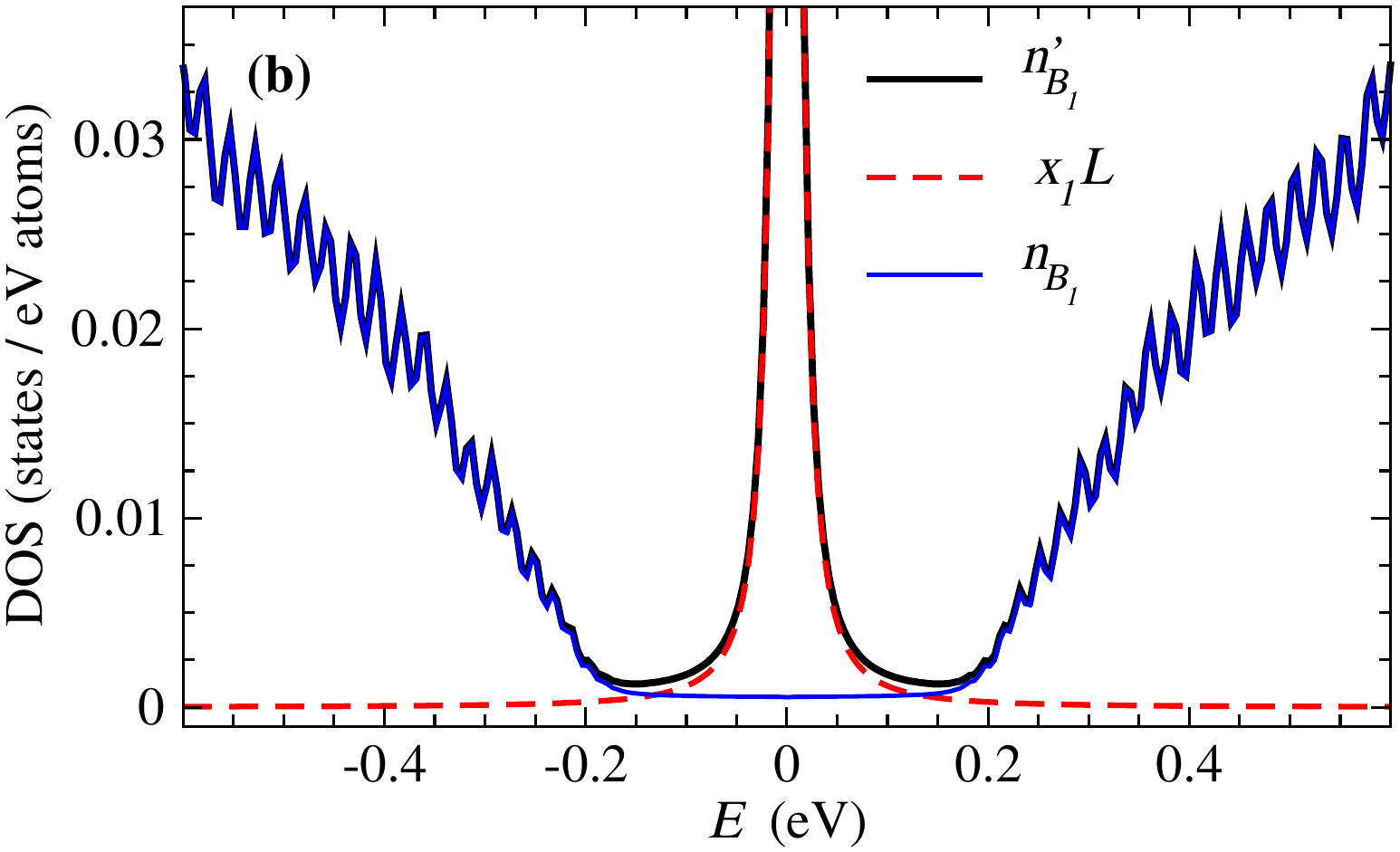} 

\caption{Electronic density of states in BLG  with B$_2$ vacant atoms ($c=0.5$\%): 
(a) Calculated total DOS $n'(E)$,  $n'(E)=n(E) + x L(E)$, with $x=0.010099$;
(b) Calculated average local DOS on A$_2$ atoms $n_{A_2}'(E)$, $n_{A_2}'(E)=n_{A_2}(E) + x_2 L(E)$, with $x_2 = 0.032573$;
(c) Calculated average local DOS on B$_1$ atoms $n_{B_1}'(E)$, $n_{B_1}'(E)=n_{B_1}(E) + x_1 L(E)$, with $x_1 = 0.0070745$.
$x_1 L(E)$ and $x_2 L(E)$ are the Lorentzian of the midgap states at $E=0$ located on B$_1$ sublattice and A$_2$ sublattice, respectively.
$n(E)$ and $n_{i}(E)$ are, respectively, the exact total DOS and the exact average local DOS on $i$ sublattice at all energies except at $E=0$. 
\label{Fig_DOS_LaB1_avec_sans_Lor}
}
\end{center}
\end{figure}

The total density of states (total DOS) $n'$ is computed by recursion (Lanczos algorithm) \cite{methodeRecursion} in real-space on sample containing $N$ carbon atoms, $N$ is up to a few $10^{7}$, with periodic boundary conditions. 
Considering a random phase states $\varphi_0$,  \cite{Roche97}
\begin{equation}
 \left | \varphi_0  \right \rangle = \frac{1}{\sqrt{N}} \sum_{m=1}^{N} {\rm e}^{i 2 \pi \theta_m}   \left | \Phi_m \right \rangle,
\end{equation}
where  $\left | \Phi_m  \right \rangle$ the $p_z$ orbital of atom $m$ 
and $\theta_m$ is a random number between 0 and 1,
the total DOS $n'$ is
\begin{equation}
n'(E) =  \lim_{\epsilon \to 0^+} n'(E,\epsilon),
\end{equation}
with
\begin{equation}
n'(E,\epsilon) = - \frac{1}{\pi} {\rm Im} \left \langle \varphi_0 \right | \frac{1}{E + i \epsilon - \hat{H}}  \left | \varphi_0  \right \rangle.
\label{eq_nepsilon}
\end{equation}
The DOS $n'$ at energy $E$ is thus evaluated numerically from a continuous fraction expansion of the Green function, $G(z) = 1 / (z-\hat{H})$, with $z = E + i \epsilon$ and a finite small $\epsilon$ value \cite{methodeRecursion}. 
Therefore, the computed DOS $n'$ is the real DOS convoluted with the Lorentzian function,
\begin{equation}
L(E,\epsilon) = \frac{\epsilon}{\pi} \frac{1}{E^2 + \epsilon^2},
\label{eq_Lorent}
\end{equation}
of half width $2 \epsilon$.
Roughly speaking, $\epsilon$ is a kind of energy precision of the calculation. As $\epsilon$ is small, $N$ should be large.
DOSs presented in this paper are computed with $\epsilon = 5$\,meV.

Average local DOSs, $n_i'$ on a $i$ sublattice, $i=$ A$_1$, B$_1$, A$_2$, B$_2$ are obtained using the same numerical method with a 
random phase $\varphi_{0i}$ expands on the $p_z$ orbitals $ \Phi_{m_i}$  of the $i$ sublattice,
\begin{equation}
 \left | \varphi_{0i}  \right \rangle = \frac{1}{\sqrt{N_i}} \sum_{m_i=1}^{N_i} {\rm e}^{i 2 \pi \theta_{m_i}}   \left | \Phi_{m_i} \right \rangle ,
\end{equation}
where $m_i$ is the index of atoms of the $i$ sublattice that contains $N_i$ atoms. 
Figure \ref{Fig_DOS_d1A1_B1} shows total DOS and average local DOSs $n'$, calculated with TB1,
for BLG with vacancies on A$_2$ and B$_2$ sublattice respectively.

As explained in the paper, with TB1 model and a $\alpha$-$\beta$ bipartite lattice, vacancies in the $\alpha$ sublattice involve midgap states, at energy $E_{MG}=0$, that are located on $\beta$ sublattice.
Because of the convolution with the Lorentzian (\ref{eq_Lorent}), the calculated total DOS $n'$ and average local DOS $n_i'$, $i=$ A$_1$, B$_1$, A$_2$, B$_2$ can be written as
\begin{equation}
n'(E,\epsilon) = n(E,\epsilon) + x L(E,\epsilon),
\label{Eq_DOS_avec_sans_Lor}
\end{equation}
and
\begin{equation}
n_i'(E,\epsilon) = n_i(E,\epsilon) + x_i L(E,\epsilon),
\label{Eq_DOS_avec_sans_Lori}
\end{equation}
where $x$ ($x_i$) is proportional of the concentration $c$ of vacant atoms, $x L$ ($x_i L$) is the calculated DOS due to midgap states, 
and $n$ ($n_i$) is the DOS (LDOS) without the midgap states.
$n'(E)$ ($n_i'(E)$) is computed by recursion method from  equation (\ref{eq_nepsilon}) with $\epsilon = 5$\,meV,
and $n(E,\epsilon)$ ($n_i(E,\epsilon)$) 
is computed from  $n'(E,\epsilon)$ ($n_i'(E,\epsilon)$) by equation ({\ref{Eq_DOS_avec_sans_Lor}) (equation ({\ref{Eq_DOS_avec_sans_Lori})).
The total DOSs $n$, shown in the main paper figures 
2(1.a) 
and 
2(1.b)
(see also figures \ref{Fig_DOS_d1A1B1_mLor}(a.1) and \ref{Fig_DOS_d1A1B1_mLor}(b.1)), 
correspond respectively to total DOSs $n'$ shown figures \ref{Fig_DOS_d1A1_B1}(a.1) and \ref{Fig_DOS_d1A1_B1}(b.1).
Showing $n$ instead of $n'$
allows to discuss more explicitly the DOS around $E_{MG} = 0$
and to show the presence of quasi-gap. This quasi-gap is hidden by the midgap state Lorentzian in DOS $n_t'$.
Figures \ref{Fig_DOS_LaA1_avec_sans_Lor} and  \ref{Fig_DOS_LaB1_avec_sans_Lor} show several examples of the three terms of equations (\ref{Eq_DOS_avec_sans_Lor})-(\ref{Eq_DOS_avec_sans_Lori}).
With A$_2$ vacant atoms (figure \ref{Fig_DOS_d1A1_B1}(a.2) and figure \ref{Fig_DOS_LaA1_avec_sans_Lor}), the midgap states 
($i.e.$ term $L(E)$ in equations (\ref{Eq_DOS_avec_sans_Lor}-\ref{Eq_DOS_avec_sans_Lori})) 
are located on B$_2$ sublattice.
With B$_2$ vacant atoms (figure \ref{Fig_DOS_d1A1_B1}(b.2) and figure \ref{Fig_DOS_LaB1_avec_sans_Lor}), 
the midgap states are located both on A$_2$ sublattice and B$_1$ sublattice.

\section*{3. Eigenstates of a square bipartite lattice with vacancies in a sublattice}

\begin{figure}
\begin{center}

\includegraphics[width=7cm]{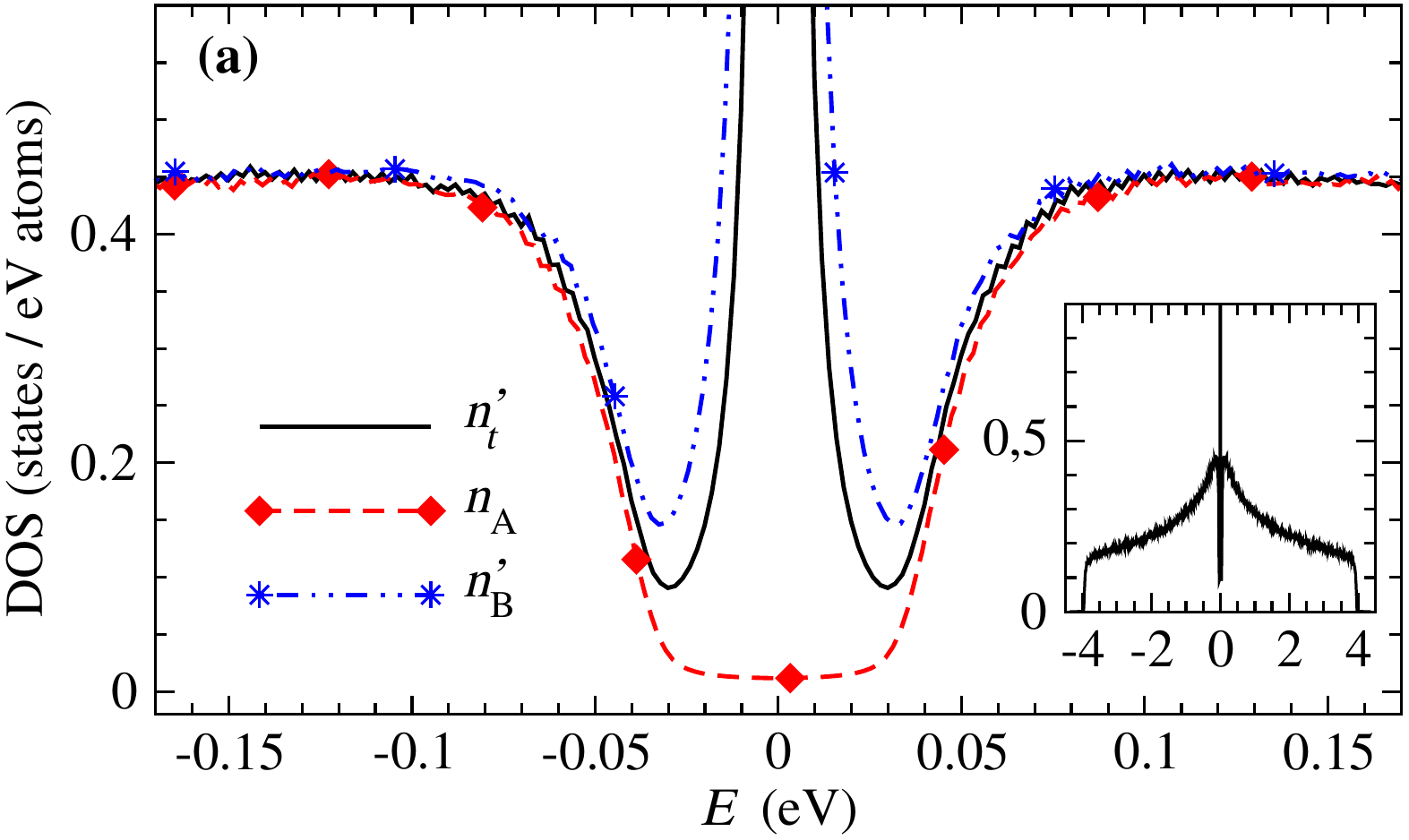} 

\includegraphics[width=7cm]{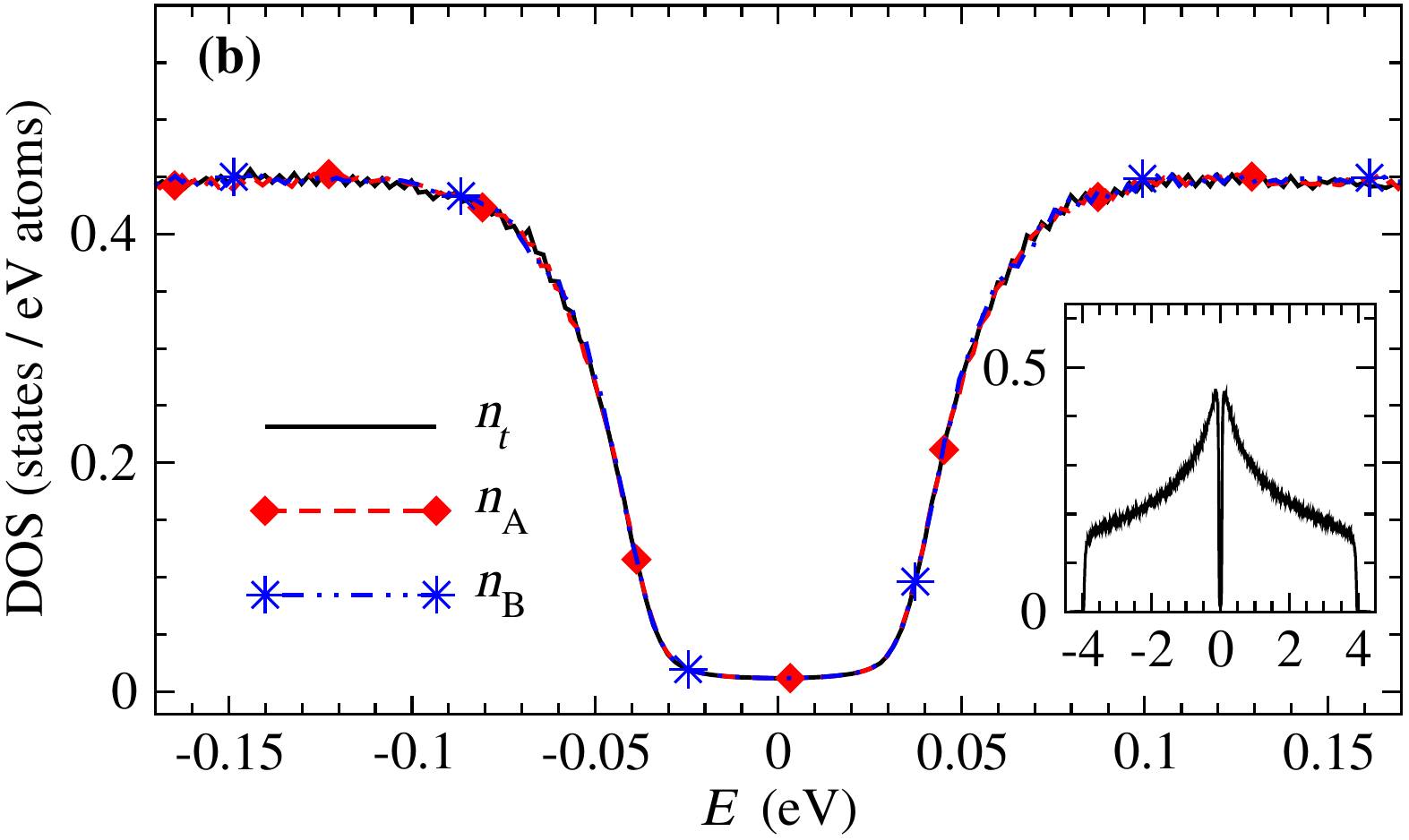} 

\includegraphics[width=7cm]{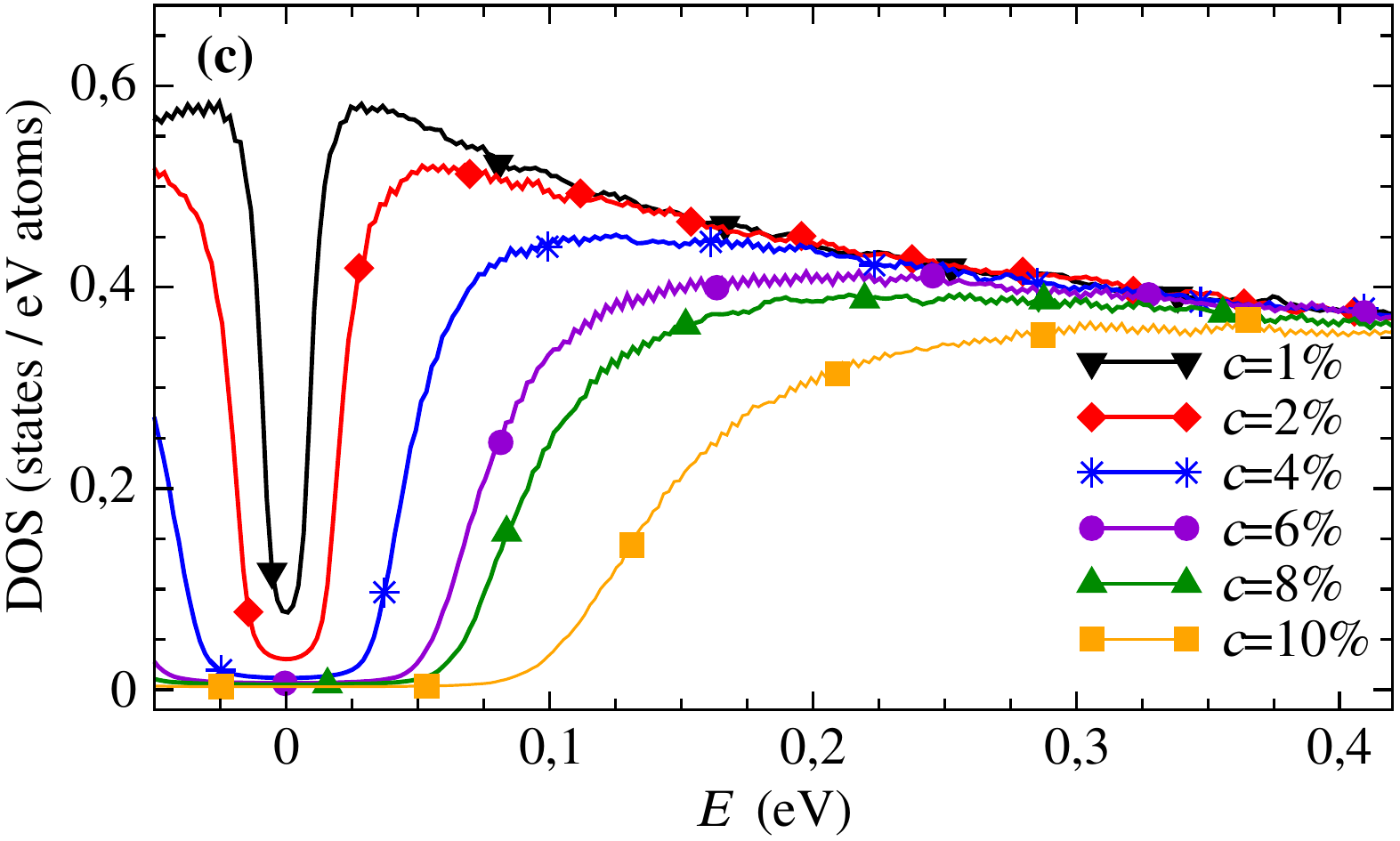} 

\includegraphics[width=7cm]{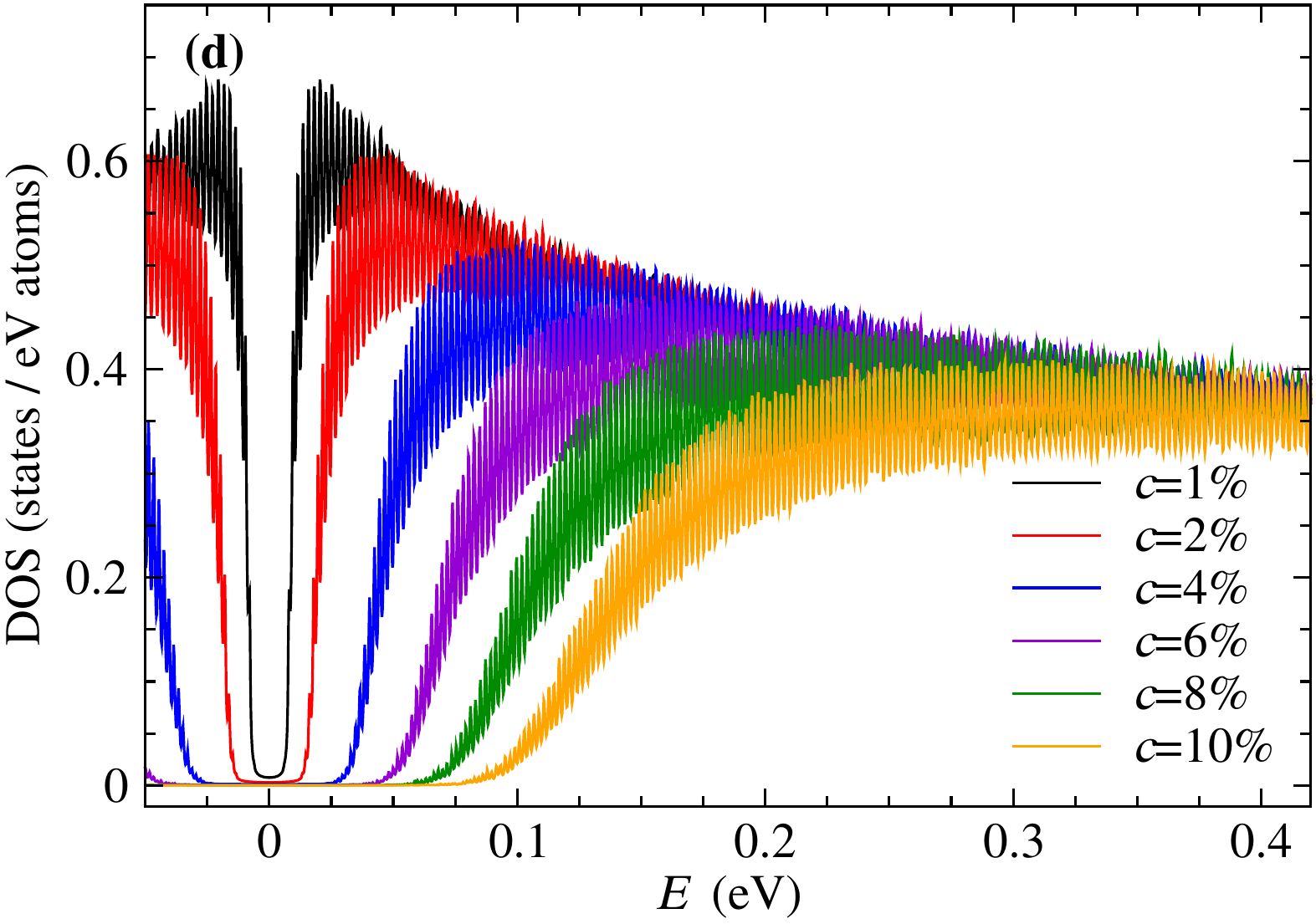}

\caption{ \label{Fig_DOS_bipartite_carre}
Bipartite square lattice with two atoms, A and B, per unit cell, and vacancies in the A (or B) sublattice:  
(a) Total DOS, $n'$, and average local DOS, $n_A' = n_A$ ($n_B'$), on A sublattice (B sublattice), for $c=4\%$ of vacant A atoms  and $\epsilon = 2$\,meV 
[Inset: total DOS $n'$].
(b) Total DOS $n$ and average local DOS $n_A$ ($n_B$) on A sublattice (B sublattice), for $c=4\%$ of vacant A atoms  and $\epsilon = 2$\,meV 
[Inset: total DOS $n$]. 
(c) and (d) Total DOS $n$ for various concentrations $c$ of vacant A atom. $c$ is the concentration of vacancies with respect to the total number of atoms in the lattice: (c) with $\epsilon = 2$\,meV and (d) $\epsilon = 0.2$\,meV.
}
\end{center}
\end{figure}

In this section, we consider a simple square lattice containing two atoms in a unit cell: atom A in position $(0,0)$ and atom B in position $(1/2, 1/2)$.
Only $s$ orbitals are taken into account and the Hamiltonian includes only nearest-neighbor hopping between orbitals:
\begin{equation}
\hat{H} = \sum_{\langle i,j \rangle} t_{ij} \left( c_i^{\dag}c_j + c_j^{\dag}c_i \right),
\label{Eq_hamiltonian_SupMat2}
\end{equation}
where $c_i^{\dag}$ and $c_i$ create and annihilate respectively an electron on the $s$ orbital located at $\vec r_i$,
$\langle i,j\rangle$ is  the sum on index $i$ and $j$ with $i\ne j$, 
and  $t_{ij}$ is the hopping matrix element
between two orbitals located at $\vec r_i$ and $\vec r_j$.
Each atom A is coupled with only B atoms and reciprocally.
This lattice is thus a bipartite lattice with two equivalent sub-parts  $\alpha$ and $\beta$  corresponding respectively to atoms A and atoms B. 

We consider  vacancies (missing atoms) distributed randomly in the A sublattice.
As explained in previous section (section 2), we calculated DOS, $n_i'$, $i = t$ (total), A, B 
(figure \ref{Fig_DOS_bipartite_carre}(a)) 
and the DOS $n_i$, $n_i(E) = n_i'(E) - x_i L(E)$, 
(figure  \ref{Fig_DOS_bipartite_carre}(b)).
The Lorentzian of the midgap states at $E=0$ is clearly seen on total DOS, $n'$,  and average local DOS, $n_B'$, on the B sublattice.
As expected from the bipartite analysis explained in the paper, vacancies in a sublattice result in a gap between  $-E_{min}$ and $E_{min}$. 
The energy gap
$2E_{min}$ increases as the concentration $c$ of vacancies increases
(figures \ref{Fig_DOS_bipartite_carre}(c,d)).

These calculations are performed by recursion in a supercell (section 2) containing $N=2 \times 2400^2 = 23.04 \times 10^6$ atoms. 
Figures \ref{Fig_DOS_bipartite_carre}(a,b,c) presents DOS calculated with $\epsilon= 2$\,meV. In this case, the convolution with the Lorentzian of width $2\epsilon$ hides the gap when the concentration of vacancies is small. To check the presence of real gap, we show DOS calculated with $\epsilon= 0.2$\,meV in figure \ref{Fig_DOS_bipartite_carre}(d). In this case, gap in clearly seen at all calculated vacancy concentrations, but unphysical oscillations appeared in DOSs. 
Those oscillations  are due to numerical artifact related to the termination of the 
continuous fraction expansion of the Green function (equation  (\ref{eq_nepsilon})) in presence of a gap \cite{methodeRecursion}.

\section*{4. Numerical method for conductivity}

\subsection*{4.1. Kubo-Greenwood scheme}

In Kubo-Greenwood formula for transport properties, 
the quantum diffusion $D$, is computed by using the polynomial expansion of the average square spreading, $\Delta X^{2}$, for charge carriers.
This method, developped by Mayou, Khanna, Roche and Triozon \cite{Mayou88,Mayou95,Roche97,Roche99,Triozon02},  
allows very efficient numerical calculations by recursion in real-space that taken into account all quantum effects.
It has been used to studies quantum transport in disordered graphene
\cite{Lherbier08,Lherbier11,Leconte11,Leconte11b,Roche12,Roche13,Trambly13} 
and chemically doped graphene \cite{Lherbier08b,Leconte10}. 
Static defects are included directly in the structural modelisation of the system and they are randomly distributed on a supercell containing up to $10^7$ Carbon atoms. 
This corresponds to typical sizes of about one micrometer square which allows to study systems with inelastic mean-free length of the order of few hundreds nanometers.
Inelastic scattering is computed \cite{Trambly13} within the
Relaxation Time Approximation. An inelastic scattering time $\tau_i$ beyond which the propagation becomes diffusive due to the destruction of coherence by inelastic process. One finally get the Einstein conductivity  formula (at $0$\,K), \cite{Trambly13}
\begin{equation}
\sigma(E_F,\tau_i) = e^2 n(E_F) D(E_F,\tau_i),
\label{eq_einstein}
\end{equation}
where $E_F$ is the Fermi level,
$D(E,\tau_i)$ is the diffusivity (diffusion coefficient at energy $E$ and inelastic scattering time $\tau_i$),
\begin{equation}
D(E,\tau_i) = \frac{L_i^2(E,\tau_i)}{2 \tau_i},
\end{equation}
$n(E)$ is the density of states (DOS) and $L_i(E,\tau_i)$ is the inelastic mean-free path. $L_i(E,\tau_i)$ is the typical distance of propagation during the time interval $\tau_i$ for electrons at energy $E$, 
\begin{equation}
L_{i}^{2}(E,\tau_{i})=\frac{1}{\tau_{i}} \int_0^\infty \! \Delta X^{2}(E,t)\,{\rm e}^{-t/\tau_{i}}  .
\end{equation}
Without static defects (static disorder) the $L_i$ and $D$ goes to infinity when $\tau_i$ diverges. 
With statics defects, at every energy $E$, $\sigma(\tau_i)$ reaches a maximum value,  
\begin{equation}
\sigma_m(E_F,\tau_i) = e^2 n(E_F) \, {\rm Max}_{\tau_i} \left \{ D(E_F,\tau_i) \right \},
\label{eq_einstein_SigM}
\end{equation}
called 
microscopic conductivity. $\sigma_m$ corresponds to the usual semi-classical approximation (semi-classical conductivity). 
This conductivity is typically the conductivity at room temperature, when inelastic scattering $\tau_i$ (inelastic mean free path $L_i$) is closed to elastic scattering $\tau_e$ (elastic mean free path $L_e$), $\tau_e(E) = L_e(E) / v(E)$ and $L_e(E) = D_m(E) / 2 v(E)$, where $D_m(E)$ is the maximum value of $D(\tau_i)$ at energy $E$ and $v(E)$ the velocity at very small times (slope of $\Delta X(t)$).

For larger $\tau_i$ and $L_i$, $\tau_e \ll \tau_i$ and $L_e \ll L_i$, quantum interferences may result in a diffusive state, 
$D(\tau_i) \simeq D_m$, or a sub-diffusive state where $D$ decreases when $\tau_i$ and $L_i$ increase. 
For very large $L_i$, $L_i$ closed to localization length $\xi$, the conductivity goes to zero. 
These two last regimes ($L_e \ll L_i$, and  $L_i \simeq \xi$), which correspond to the low temperature regime,  are not discussed in the main paper.

\begin{figure}
\begin{center}

\includegraphics[width=4.2cm]{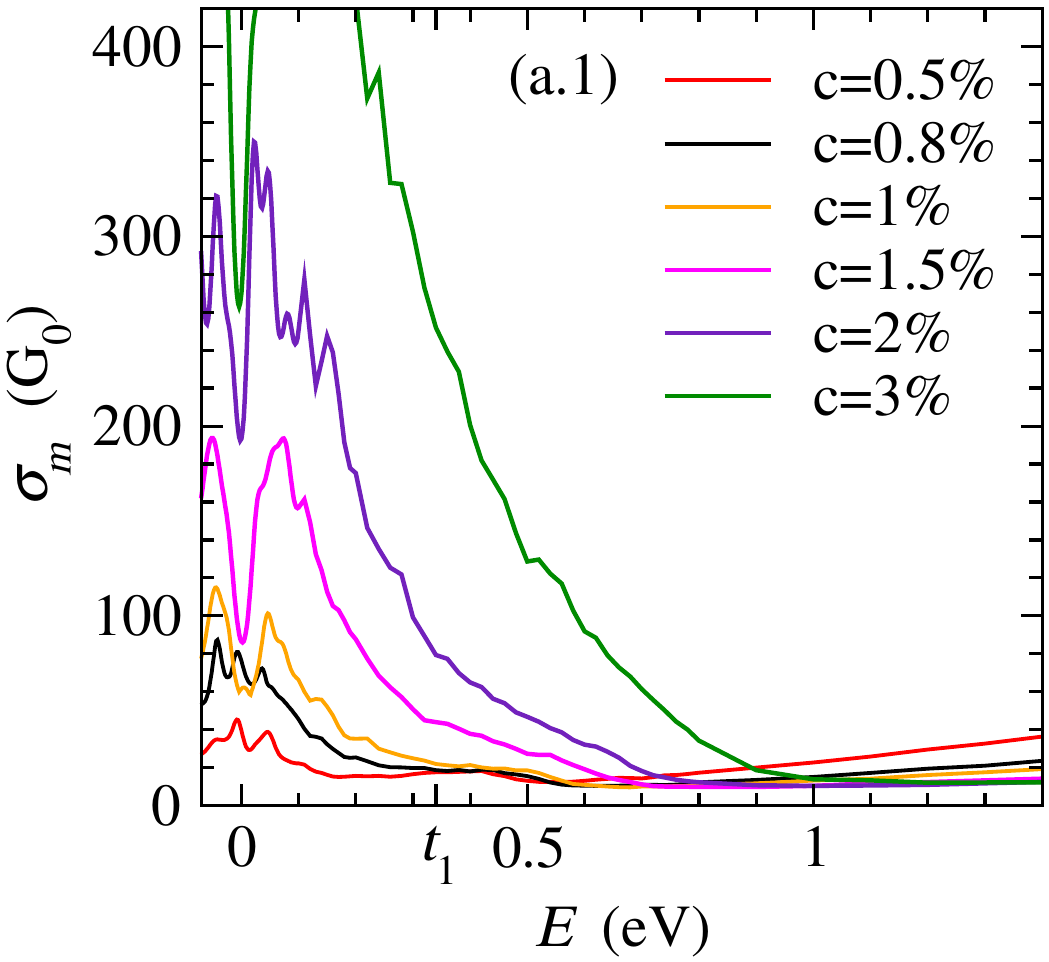} 
~ \includegraphics[width=4.1cm]{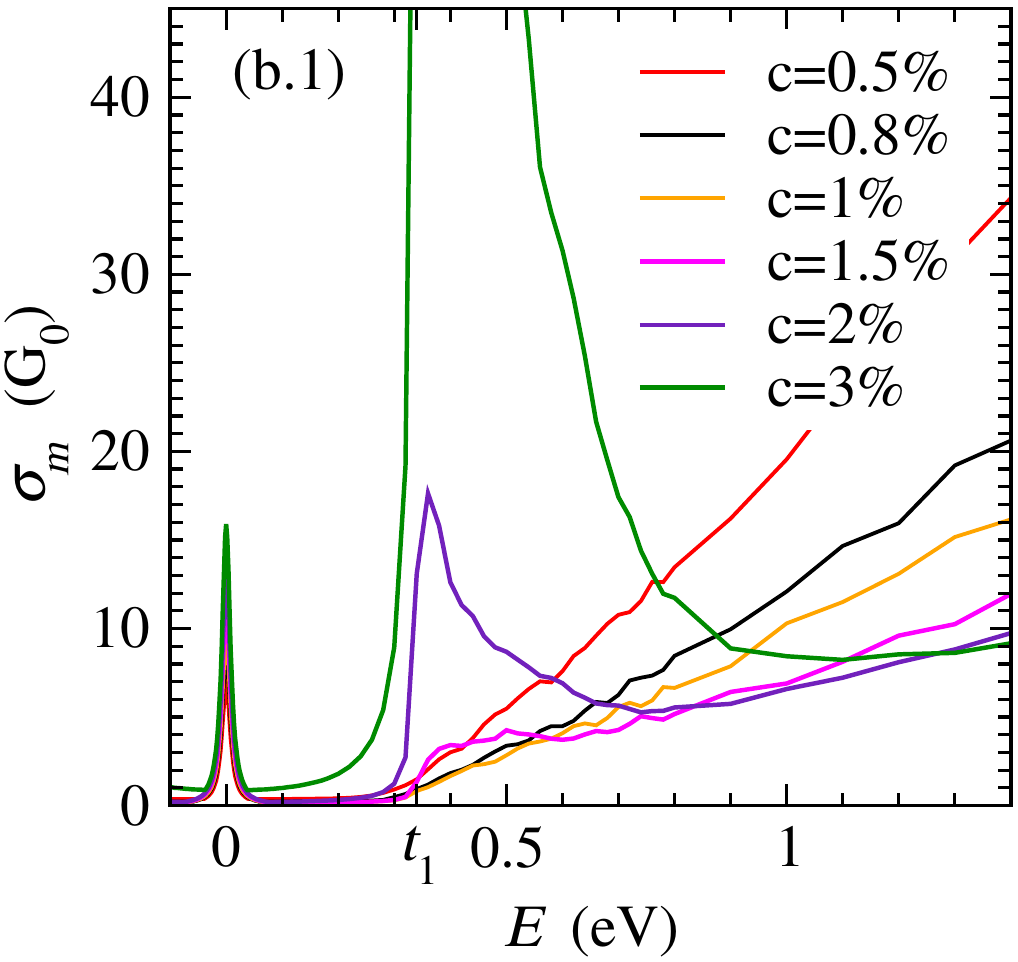} 

\vspace{.1cm}
\includegraphics[width=4.1cm]{Fig_d1VaA1_SigM_mLorDOS} 
~ \includegraphics[width=4.1cm]{Fig_d1VaB1_SigM_mLorDOS}

\caption{ \label{Fig_DOS_Sig_d1A1B1_mLorDOS}
Microscopic conductivity  $\sigma_m(E)$ computed from TB1 in BLG: 
for A$_2$ vacant atoms [(a.1) (a.2)], 
and for B$_2$ vacant atoms [(b.1) (b.2)].
$\sigma_m(E)$ is computed from equation (\ref{eq_einstein}) 
where (a.1) (b.1) the total DOS $n(E)$ includes Lorentzian due to midgap states at $E=0$, 
and (a.2) (b.2) $n(E)$ excludes the Lorentzian due to midgap states at $E=0$.
$c$ is the concentration of vacancies with respect to the total number of atom in BLG.
Spectrum is symmetric with respect to Dirac energy $E_D=0$.
}
\end{center}
\end{figure}

\subsection*{
4.2. Midgap states in case of first neighbor hopping model (TB1) with vacancies on a sublattice only}

As discussed section 2, with first neighbor hopping model (TB1) and vacancies in only one sublattice, midgap states are found at the energy $E_D = 0$. These states are not coupled to each other by $H$; thus they do not contributed to the DOS and the conduction properties at $E \ne 0$. 
For this reason, the conductivity presented in the main paper is computed numerically by equation (\ref{eq_einstein_SigM}) using the total DOS $n(E)$ that does not included the Lorentzian due to midgap states at $E = 0$ (section 2 of this Supplementary Material). 
Microscopic conductivity computed with and without this Lorentzian contribution to the DOS is shown figure \ref{Fig_DOS_Sig_d1A1B1_mLorDOS}.

Therefore our study does not deal with conduction by midgap states at $E \ne 0$ in the case of first neighbor hopping model (TB1).
Nevertheless the results with the more realistic Hamiltonian TB2, that includes hopping beyond first neighbor, show that midgap states in TB1 model are very specific to TB1 model and are not realistic in real bilayer graphene, as it has been found for the monolayer graphene \cite{Trambly14}.

\section*{References}

\bibliography{biblio_biAB}
\bibliographystyle{unsrt}

\end{document}